\documentclass[twoside,twocolumn,9pt]{article}
\usepackage{extsizes}
\usepackage[super,sort&compress,comma]{natbib} 
\usepackage[version=3]{mhchem}
\usepackage[left=1.5cm, right=1.5cm, top=1.785cm, bottom=2.0cm]{geometry}
\usepackage{balance}
\usepackage{mathptmx}
\usepackage{sectsty}
\usepackage{graphicx} 
\usepackage{lastpage}
\usepackage[format=plain,justification=justified,singlelinecheck=false,font={stretch=1.125,small,sf},labelfont=bf,labelsep=space]{caption}
\usepackage{float}
\usepackage{fancyhdr}
\usepackage{fnpos}
\usepackage[english]{babel}
\addto{\captionsenglish}{%
  \renewcommand{\refname}{Notes and references}
}
\usepackage{array}
\usepackage{droidsans}
\usepackage{charter}
\usepackage[T1]{fontenc}
\usepackage[usenames,dvipsnames]{xcolor}
\usepackage{setspace}
\usepackage[compact]{titlesec}
\usepackage{hyperref}

\usepackage{epstopdf}

\definecolor{cream}{RGB}{222,217,201}

\begin{document}

\pagestyle{fancy}
\thispagestyle{plain}
\fancypagestyle{plain}{
\renewcommand{\headrulewidth}{0pt}
}

\makeFNbottom
\makeatletter
\renewcommand\LARGE{\@setfontsize\LARGE{15pt}{17}}
\renewcommand\Large{\@setfontsize\Large{12pt}{14}}
\renewcommand\large{\@setfontsize\large{10pt}{12}}
\renewcommand\footnotesize{\@setfontsize\footnotesize{7pt}{10}}
\makeatother

\renewcommand{\thefootnote}{\fnsymbol{footnote}}
\renewcommand\footnoterule{\vspace*{1pt}%
\color{cream}\hrule width 3.5in height 0.4pt \color{black}\vspace*{5pt}} 
\setcounter{secnumdepth}{5}

\makeatletter 
\renewcommand\@biblabel[1]{#1}            
\renewcommand\@makefntext[1]%
{\noindent\makebox[0pt][r]{\@thefnmark\,}#1}
\makeatother 
\renewcommand{\figurename}{\small{Fig.}~}
\sectionfont{\sffamily\Large}
\subsectionfont{\normalsize}
\subsubsectionfont{\bf}
\setstretch{1.125} 
\setlength{\skip\footins}{0.8cm}
\setlength{\footnotesep}{0.25cm}
\setlength{\jot}{10pt}
\titlespacing*{\section}{0pt}{4pt}{4pt}
\titlespacing*{\subsection}{0pt}{15pt}{1pt}

\fancyfoot{}
\fancyfoot[LO,RE]{\vspace{-7.1pt}\includegraphics[height=9pt]{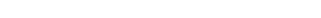}}
\fancyfoot[CO]{\vspace{-7.1pt}\hspace{11.9cm}\includegraphics{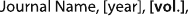}}
\fancyfoot[CE]{\vspace{-7.2pt}\hspace{-13.2cm}\includegraphics{RF}}
\fancyfoot[RO]{\footnotesize{\sffamily{1--\pageref{LastPage} ~\textbar  \hspace{2pt}\thepage}}}
\fancyfoot[LE]{\footnotesize{\sffamily{\thepage~\textbar\hspace{4.65cm} 1--\pageref{LastPage}}}}
\fancyhead{}
\renewcommand{\headrulewidth}{0pt} 
\renewcommand{\footrulewidth}{0pt}
\setlength{\arrayrulewidth}{1pt}
\setlength{\columnsep}{6.5mm}
\setlength\bibsep{1pt}

\makeatletter 
\newlength{\figrulesep} 
\setlength{\figrulesep}{0.5\textfloatsep} 

\newcommand{\topfigrule}{\vspace*{-1pt}%
\noindent{\color{cream}\rule[-\figrulesep]{\columnwidth}{1.5pt}} }

\newcommand{\botfigrule}{\vspace*{-2pt}%
\noindent{\color{cream}\rule[\figrulesep]{\columnwidth}{1.5pt}} }

\newcommand{\dblfigrule}{\vspace*{-1pt}%
\noindent{\color{cream}\rule[-\figrulesep]{\textwidth}{1.5pt}} }

\makeatother

\twocolumn[
  \begin{@twocolumnfalse}
{\includegraphics[height=30pt]{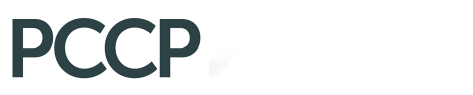}\hfill\raisebox{0pt}[0pt][0pt]{\includegraphics[height=55pt]{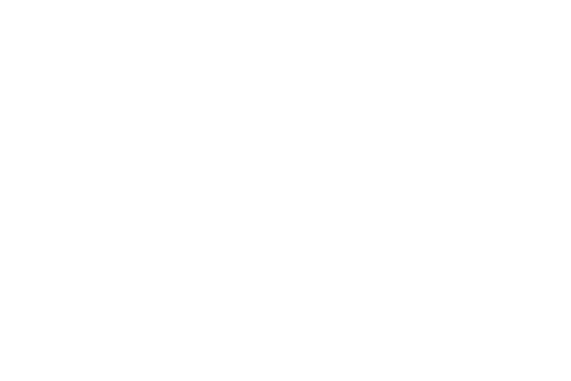}}\\[1ex]
\includegraphics[width=18.5cm]{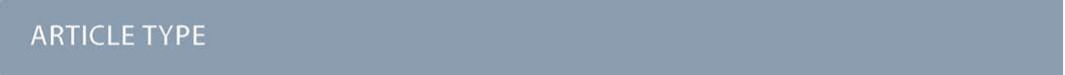}}\par
\vspace{1em}
\sffamily
\begin{tabular}{m{4.5cm} p{13.5cm} }

\includegraphics{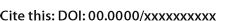} & \noindent\LARGE{\textbf{Ultrafast Processes in 1,2-Dichloroethene measured with a Universal XUV probe$^\dag$}} \\
\vspace{0.3cm} & \vspace{0.3cm} \\

 & \noindent\large{Henry G. McGhee,\textit{$^{a}$} Henry J. Thompson,\textit{$^{b}$} James Thompson,\textit{$^{c}$} Yu Zhang,\textit{$^{c}$} Adam S. Wyatt,\textit{$^{c}$} Emma Springate,\textit{$^{c}$} Richard T. Chapman,\textit{$^{c}$} Daniel A. Horke,\textit{$^{d}$}, Russell S. Minns,\textit{$^{b}$} Rebecca A. Ingle,\textit{$^{a}$} and Michael A. Parkes$^{\ast}$\textit{$^{a}$}}\\

\includegraphics{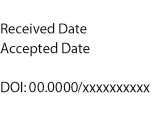} & \noindent\normalsize{The presence of two chlorine atoms in 1,2-dichloroethene allows for isomerisation around the double bond. This isomerisation can lead to rich photochemistry. We present a time-resolved pump-probe photoelectron spectroscopy measurement on both the \emph{cis}- and \emph{trans}- isomers of 1,2-dichloroethene. A universal XUV probe of 22.3~eV is used allowing observation of photoelectrons formed anywhere on the potential energy surface, even from the ground-state or dissociation products. Following excitation with a 200~nm probe both ultrafast excited state dynamics and product formation are observed within the time resolution of the experiment. Excited state population begins to return to the ground state on an ultrafast time scale (\textless{} 70~fs) and population of products channels is observed on the same timescale. With the aid of \emph{ab initio} calculations it is found that population transfer from the excited state is facilitated by vibrational modes involving C-C-H bends.
}
\end{tabular}

 \end{@twocolumnfalse} \vspace{0.6cm}

  ]

\renewcommand*\rmdefault{bch}\normalfont\upshape
\rmfamily
\section*{}
\vspace{-1cm}

\footnotetext{\textit{$^{a}$~Department of Chemistry, University College London, 20 Gordon Street, London, WC1H 0AJ, UK. E-mail: michael.parkes@ucl.ac.uk}}
\footnotetext{\textit{$^{b}$~School of Chemistry, University of Southampton, Highfield, Southampton SO17 1BJ, UK. }}
\footnotetext{\textit{$^{c}$~Central Laser Facility, STFC Rutherford Appleton Laboratory, Didcot, Oxfordshire OX11 0QX, UK}}
\footnotetext{\textit{$^{d}$~Radboud University, Institute for Molecules and Materials, Heijendaalseweg 135, 6525 AJ Nijmegen, The Netherlands }}

\footnotetext{\dag~Electronic Supplementary Information (ESI) available: [Calculated geometries and vibrational frequencies for the molecules. Further experimental results for \emph{cis}-1,2-dichloroethene]. See DOI: 10.1039/cXCP00000x/}


\section{Introduction}

Following the absorption of a photon, the excess internal energy of a molecule can be redistributed \emph{via} several competing photophysical and photochemical processes. These processes, which often occur on ultrafast timescales, underpin many fundamental mechanisms found in nature. The C=C double bond, a key component in organic molecules, is of particular interest in this context. This chemical motif exhibits rich photochemistry, with isomerisation around the double bond playing a crucial role in the photochemical pathway of vision.\cite{retinalTR} 
 
\begin{figure}
    \centering
    \includegraphics[width=5cm]{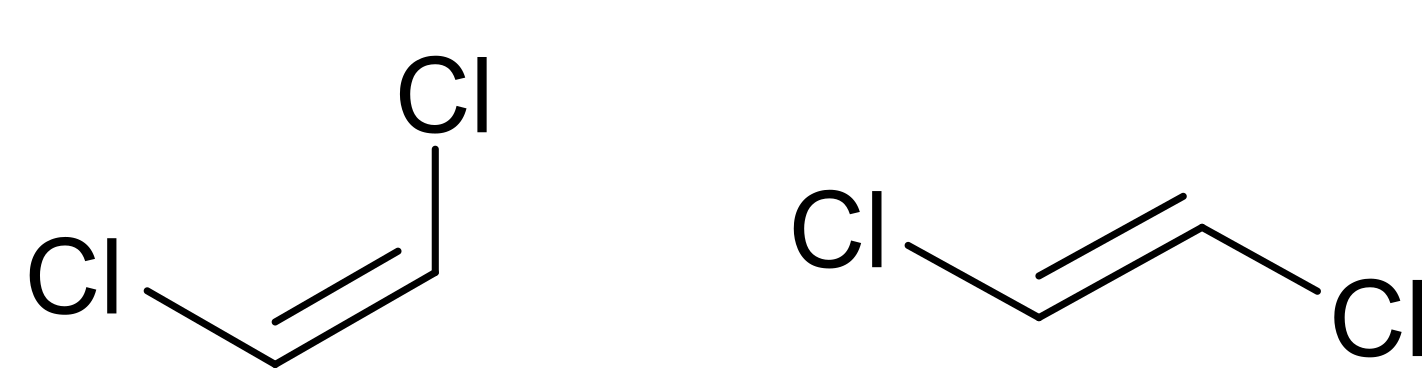}
    \caption{Structures of \emph{cis} and \emph{trans}-1,2-dichloroethene}
    \label{fig:structures}
\end{figure}

The \emph{cis} and \emph{trans} isomers of 1,2-dichloroethene (DCE) are examples of substituted ethenes, in which two H atoms are replaced with Cl atoms (Fig. \ref{fig:structures}). Whereas many ethylenic systems are known to photoisomerise on exposure to UV light,\cite{Karashima2022,wang2022stilbene} the inclusion of the reasonably weak C-Cl bonds opens up the possibility of additional photoinduced pathways, including photodisocciation to form a range of products including Cl radicals. The DCEs have therefore attracted a great deal of interest experimentally and theoretically in the past decades, both from the stand point of fundamental photochemistry and in the context of atmospheric research.\cite{atmos_chloros,OH-Yamada2001,OH-CanosaMas2001} 


DCEs show a broad resonance in their UV absorption spectrum with a maximum at $\sim$200~nm for \emph{trans} and $\sim$190~nm for \emph{cis}. Previous vertical excitation energy calculations at the TDDFT/aug-cc-pVDZ level of theory suggest that the state with the majority of the oscillator strength is a $\pi\pi^*$ transition located on the C=C double bond, although there are a number of weak Rydberg series present in the same energetic region.\cite{Locht2019-cis,Locht2020-trans} There are also dissociative $n\sigma^*$ and $\pi\sigma^*$ states that are energetically accessible, but have almost zero oscillator strength at the Franck-Condon geometry.

Early photolysis studies revealed photodissociation products for both isomers at $\sim$200~nm as \ce{C2H2}, \ce{C2HCl}, \ce{HCl}, \ce{C2H2Cl^.}, and \ce{Cl^.}.\cite{Wijnen1961,Ausubel1975trans,Ausubel1975cis,Berry1974,Moss1981} By measuring photofragment translational energy distributions, Umemoto \emph{et al.} discovered that Cl is produced with two distinct energy releases implying the presence of two dissociation channels: one with a Gaussian velocity distribution showing angular anisotropy and another with a Boltzmann-like velocity distribution and weak angular anisotropy.\cite{Umemoto1985_Cl_dis} The former is ascribed to direct crossing of the $\pi\pi^*$ state with nearby $n\sigma^*$ and $\pi\sigma^*$ states, and the latter from the dissociation of vibrationally excited ground state molecules following internal conversion. Subsequent studies support this interpretation and further showed the branching ratio of these channels to exhibit a wavelength dependence.\cite{Mo1992,Sato1993,Suzuki1994_ionimage,Huang1995,Hua2010} In contrast, molecular elimination to produce HCl and C\textsubscript{2}HCl is believed to occur solely on the electronic ground state with a higher yield for the \emph{trans} isomer.\cite{He1993} 

However, these reaction schemes fail to address the presence of acetylene (C\textsubscript{2}H\textsubscript{2}) as a photoproduct. The absence of its dissociative partner Cl\textsubscript{2} led to the proposal that acetylene is formed from the secondary dissociation of the vibrationally excited chlorovinyl radical (following primary dissociation of the initial Cl atom).\cite{Sato1997,Seki2011} This mechanism also explains the presence of two Cl channels for which the branching ratio exhibits a wavelength dependence. Other potential pathways involving H and H\textsubscript{2} elimination have been shown to have very low yield in this wavelength range.\cite{Mo1992,He1993}

The ambiguities in the dissociation mechanism along with the absence of timescales for these processes prompted us to probe deeper into the photoinduced dynamics of DCEs. Time-resolved Photoelectron Spectroscopy (TRPES) provides a highly effective way of probing the chemical dynamics of the system\cite{Stolow2004} to help understand the photodissociation processes that occur in DCE. By using an XUV probe, there is sufficient probe energy to ionize all species on the reaction pathway, making it sensitive to large geometrical changes from the ground state. It is therefore an appropriate tool to follow photodissociation dynamics in which multiple fragments with a range of geometries may be produced. TRPES using an XUV source has been shown to be successful in the probing of chemical dynamics using both HHG sources\cite{Karashima2022,Smith:PRL120:183003,wang2022stilbene} and XFELS.\cite{Pathak2020,Borne2024}
 
Here we apply TRPES to \emph{cis} and \emph{trans} dichloroethene using a 200~nm pump and a universal XUV probe. The experimental results are presented alongside \emph{ab initio} calculations of the molecules' potential energy surfaces. The kinetics of the observed excited state features and photofragments are discussed with regard to prior studies, and serve to further improve our understanding of the dynamics of these photochemically interesting molecules.

    \section{Experimental}\label{Experimental}
    
        Experiments were performed using the AMO end station at the Artemis facility (Central Laser Facilities, Rutherford Appleton Laboratories). This experiment has been described in detail before so only a brief overview will be given here.\cite{Smith:PRL120:183003} \emph{cis}- and \emph{trans}-1,2-dichloroethene were purchased from Sigma Aldrich with 97\% and 98\% purity, respectively. They were used without further purification, but degassed with a cycle of freeze-pump-thawing before use. The samples entered the interaction chamber of the end station as an effusive beam with chamber pressure maintained at 1x10$^{-6}$~mbar during measurements. The inlet line was gently heated ($\sim$40~$^{\circ}$) to aid pressure stability and lower the risk of the nozzle blocking. The effusive beam was overlapped with the pump and probe laser beams at the entrance to an electron time-of-flight (eTOF) spectrometer (Kaesdorf ETF11). Any photoelectrons produced by interactions of the lasers and samples were collected in this eTOF. 
    
        The pump and probe beams were generated from a Ti:Sapphire amplifier system with a 1~kHz repetition rate (Red Dragon, KM Labs). The 200~nm pump pulse was generated from the fundamental of the Red Dragon through a series of BBO crystals. Around 1~$\mu$J of pump with a bandwidth of around 1.8~nm and pulse durations of approximately 50~fs (FWHM) was generated. The high-harmonic generation was performed by focusing the second harmonic (around 400~nm) of the fundamental into an Argon jet, the 7th harmonic was selected (22.3~eV, 57~nm) using a time-preserving monochromator.\cite{Art_mono} Probe pulses with a bandwidth of 500~meV and pulse durations of 35~fs (FWHM) was produced. The two beams (the pump and probe) were focused and recombined in front of the entrance to the eTOF. Variable time delays between the pump and the probe pulses were achieved using a mechanical delay stage.

        Both \emph{cis} and \emph{trans}-1,2-dichloroethene have $\pi\rightarrow\pi^*$ transitions around 200~nm,\cite{Locht2019-cis,Locht2020-trans} though with different labels (S$_3$ and S$_2$ respectively) due to the different ordering of the excited states. For the \emph{trans} isomer the pump wavelength corresponds to the maximum of the peak in the photoabsorption spectrum. In contrast, for the \emph{cis} isomer the pump wavelength is approximately 10~nm below the peak in the photoabsorption spectrum. However, given the large differences in oscillator strengths between the \emph{cis} transitions, we expect most of the excited state population will be in the S$_3$ state. Therefore, for both \emph{cis} and \emph{trans} the observed time-dependent dynamics will begin in the $\pi\pi^*$ state.
        
        The eTOF used had a variable energy transmission which could be tuned using a lens at the TOF entrance. For these experiments this lens was tuned to maximise transmission of electrons whose kinetic energy was close to the energy of ionization from the neutral ground state into the first ion state. This means that electrons ejected with lower energies would not be collected as efficiently. Electron signals from the eTOF were discriminated and then recorded with a time-to-digital converter (Surface Concepts GmbH - SC-TDC-1000/04 D).
    
        Pump-probe spectra were collected at a range of time delays and saved as `cycles' (each cycle = 20 sweeps of a delay sequence). There were 97 cycles for the \emph{trans} data which amounted to 84 x $10^{6}$ photoelectrons detected across all kinetic energies and time bins. For \emph{cis}, the data was recorded for 80 cycles (65 x $10^6$ electrons). A standard bootstrapping analysis was performed to estimate the statistical uncertainty. Cycles were randomly sampled with replacement 100 times and a time-of-flight to energy conversion (with Jacobian correction) was applied to each new dataset. The mean and standard deviation of these 100 datasets define the values and error bars shown in Figures \ref{fig:trans_results}, \ref{fig:trans}, \ref{fig:cis_results}, \ref{fig:cis}.

    \section{Theoretical}

        The ground state minimum energy geometries of \emph{cis} and \emph{trans}-1,2-dichloroethene were optimised at the M{\"o}ller Plesset Perturbation (MP2) level of theory with a 6-311G++(3df, 3pd) basis set in Gaussian 16.\cite{g16} 
        Vertical excitation (EOM-EE-CCSD) and ionization energies (EOM-IP-CCSD) were calculated using the optimised minimum energy structures in QChem 5.4 with a 6-311G++(3df, 3pd) basis set.\cite{qChem} Cuts of the potential energy surface (PES) were made along the normal modes and the energies of the first eight singlet excited states, all the cuts are shown in the supporting information and cuts along key modes are shown in Figure \ref{fig:key_cuts} for \emph{trans}-1,2-dichloroethene. It should be noted that CCSD methods are known to not produce accurate results in regions where the electronic wavefunction is not well described by a single configuration. Therefore, EOM-EE-CCSD and EOM-IP-CCSD calculations were not performed in those regions where the wave function is strongly multi-configurational. 
       
        Results from the EOM-CCSD calculations for both isomers are summarised in Tables \ref{tab_trans_eng} and \ref{tab_cis_eng} for \emph{trans} and \emph{cis} respectively. Comparison of the calculated excitation and ionization energies to experimental results are given for both isomers in the SI (Figures SI S2 - S5). As can be seen the calculated values agree reasonably well, the excitation energies are around 0.4~eV too high in energy (Figures S4 and S5), while the ionization energies are around 0.1~eV different (Figures S2 and S3). This close agreement suggests that the calculated results should give a reasonable representation of the potential energy surfaces. For both isomers, the theoretical results show the  $\pi\rightarrow\pi^*$ has the strongest oscillator strength of the transitions expected to be energetically accessible following 200 nm excitation. At the Franck-Condon geometry, the $\pi\rightarrow\pi^*$ corresponds to the S$_2$ state for \emph{trans}-DCE, whereas for \emph{cis}-DCE, the $\pi\rightarrow\pi^*$ corresponds to S$_3$.
        
        \begin{table}[h]
            \centering
            \caption{Calculated excited state properties for \emph{trans}-1,2-dichloroethene from an EOM-CCSD calculation with a 6-311G++(3df, 3pd) basis. For each state the label, symmetry, energy, oscillator strength (f) and character of the transition is given.}
            \begin{tabular}{lllll}
                \hline
                State & Sym. & Energy /& f & Character \\ 
                & & eV& & \\\hline
                S$_1$ & B$_g$ & 6.32 & 0 & $\pi\rightarrow\sigma^*$\textsubscript{C-Cl} + Ryd \\ 
                S$_2$ & B$_u$ & 6.66 & 0.34 & $\pi\rightarrow\pi^*$ \\ 
                S$_3$ & A$_u$ & 6.76 & 0.002 & $\pi\rightarrow$ Ryd\\ 
                S$_4$ & B$_g$ & 7.37 &  0    & $\pi\rightarrow\sigma$\textsubscript{C-H} + n\textsubscript{Cl} + Ryd\\ \hline
            \end{tabular}
            \label{tab_trans_eng}
        \end{table}
        
        \begin{table}[!ht]
            \centering
            \caption{Calculated excited state properties for \emph{cis}-1,2-dichloroethene from an EOM-CCSD calculation with a 6-311G++(3df, 3pd) basis. For each state the label, symmetry, energy, oscillator strength (f) and character of the transition is given.}
            \begin{tabular}{lllll}
            \hline
                State & Sym. & Energy /& f & Character \\ 
                 & & eV& & \\\hline
                S$_1$ & B$_1$ & 6.47 & 0.001 & $\pi\rightarrow\sigma^*$\textsubscript{C-Cl} + Ryd \\ 
                S$_2$ & B$_1$ & 6.72 & 0.001 & $\pi\rightarrow$ n\textsubscript{Cl} + Ryd\\ 
                S$_3$ & B$_2$ & 6.90& 0.36 & $\pi\rightarrow\pi^*$ \\ 
                S$_4$ & A$_2$   & 7.42  & 0 & $\pi\rightarrow\sigma$\textsubscript{C-H} + n\textsubscript{Cl} + Ryd\\ \hline
            \end{tabular}
            \label{tab_cis_eng}
        \end{table}

    \section{Results}

        \begin{figure}[h!]
            \centering
            \includegraphics{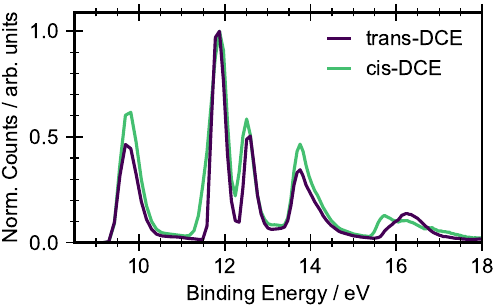}
            \caption{Ground state photoelectron spectra for the \emph{cis} and \emph{trans} isomers of dichloroethene (DCE) with an ionizing photon energy of 22.3~eV, recorded by averaging the photoelectron spectra at negative time delays (-1.5~ps to -0.4~ ps). The spectra have been normalised to the maximum of the $\tilde{\textrm{A}}$ state.}
            \label{fig:gs}
        \end{figure}
        
        Figure \ref{fig:gs} shows the ground state photoelectron spectra for \emph{cis}- and \emph{trans}-1,2-dichloroethene recorded with an incident photon energy of 22.3~eV. The photoelectron spectra of the two isomers show similar structure with an onset in signal around 9.5~eV due to ionization into the $\tilde{\textrm{X}}$ state of the cation. The clearest difference between the \emph{cis} and \emph{trans} isomers is the shape and position of the state around 16~eV. This peak is a commbination of ionization into two different ion states, $\tilde{\textrm{E}}$ and $\tilde{\textrm{F}}$. Both states arise due to ionization from orbitals that are combinations of C-Cl, C=C and C-H bonding. In the \emph{cis} isomer these orbitals have the same symmetry and are therefore non-degenerate and resolvable. However, the orbitals have different symmetries in the \emph{trans} isomer and hence have a near degeneracy. Other differences are in the relative heights of the $\tilde{\textrm{X}}$ and $\tilde{\textrm{A}}$ state peaks, and the widths of the $\tilde{\textrm{A}}$ and $\tilde{\textrm{B}}$ state peaks. The $\tilde{\textrm{A}}$ peak for \emph{cis} is much wider than for \emph{trans}. Given the onset of ionization at 9.5~eV, any signal observed with a binding energy below this will be due to ionization of a UV-excited state.

        \subsection{\emph{Trans}-1,2-dichloroethene}

        \begin{figure*}
            \centering
            \includegraphics{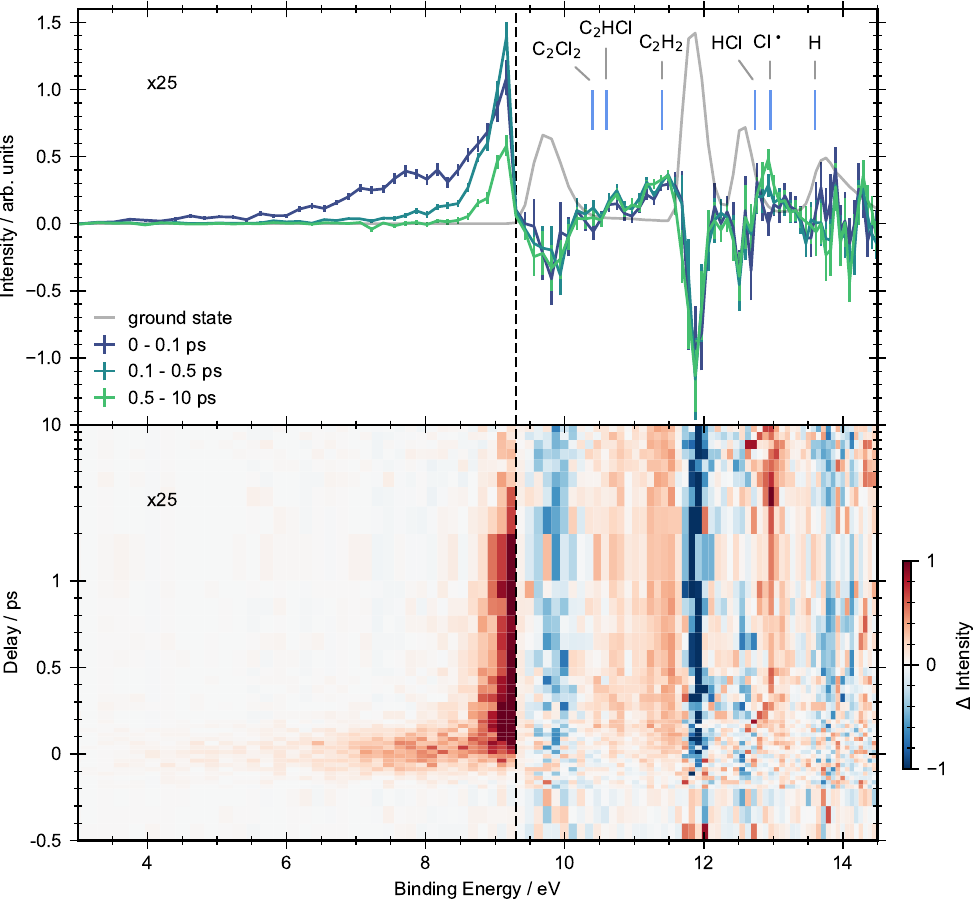}
            \caption{Pump-probe photoelectron spectra for \emph{trans}-dichloroethene with 200~nm pump and 22.3~eV probe energies.  Top panel: Grey line shows the ground-state photoelectron spectra obtained by averaging the pump-probe data at negative time delays (-1.5, -1, -0.5 and -0.4~ps). Solid lines represent the background subtracted photoelectron spectra integrated between selected pump/probe delay intervals. The error bars were obtained by standard bootstrapping analysis as described in Section \ref{Experimental}. Blue vertical lines mark the experimental vertical ionization energies of photofragments that may be produced \emph{via} photodissociation. Bottom panel: Contour plot showing the background subtracted photoelectron spectra across all pump-probe delay times. Delays greater than 1~ps are plotted on a logarithmic scale. In both panels, to aid visual interpretation the excited state signals that are observed below the onset of the first ionization band (9.3~eV, vertical dashed line) have been scaled up by a factor of $\times$25.}
             \label{fig:trans_results}
        \end{figure*}
        
        In Figure \ref{fig:trans_results} we present a heat map of the time-dependent photoelectron spectra for \emph{trans}-1,2-dichloroethene obtained following excitation with a 200~nm pump. Here we also show time-averaged photoelectron spectra over selected time windows to aid interpretation. Due to the energetic position of excited state features below the onset of ground state ionization (3.0 - 9.2~eV), they have an excellent signal-to-noise ratio, hence our analysis will focus on this energy range before considering higher binding energies.

        Upon photoexcitation, the photoelectron spectrum of trans-DCE shows the appearance of a broad feature spanning binding energies between 3.0 - 8.0 eV. While none of these features are fully resolved in the present experiment experiment, there appears to be a peak in the signal intensity at 7.6 eV and a broader feature close to 8.5 eV, suggesting that the low binding energy region of the spectrum has contributions from multiple cation states.

        The positive feature with an onset at around 8~eV shows different dynamics to the low binding energy features. This feature forms rapidly and then decays on a longer timescale with two time constants, becoming narrower in the process. A small amount of signal remains at the edge of the experimental time window (10~ps).
        
        Above the first ionization energy of dichloroethene the signal-to-noise ratio of the data becomes poorer. This is in part due to less efficient detector transmission at lower electron kinetic energies, and a strong overlapping photoionization signal from ground-state dichloroethene. However, it is possible to see time resolved signal at these higher binding energies in regions where the ground-state signal is weak. In the spectrum associated with asymptotically long pump-probe delays (0.5 - 10~ps) clear signals in the binding energy regions associated with the dissociation products (C$_2$Cl$_2$ (10.40~eV), C$_2$HCl (10.60~eV), C$_2$H$_2$ (11.40~eV) and Cl$^.$ (12.96~eV)) are observed. No signals associated with the HCl (12.74~eV) or H (13.60~eV) products are observed, probably due to their overlap with a large ground state background. Ground-state bleach signals are observed for all the bands in the ground-state spectrum, none of which undergo full recovery on the timescale of the experiment. 
         
        \begin{figure}[h!]
            \centering
            \includegraphics{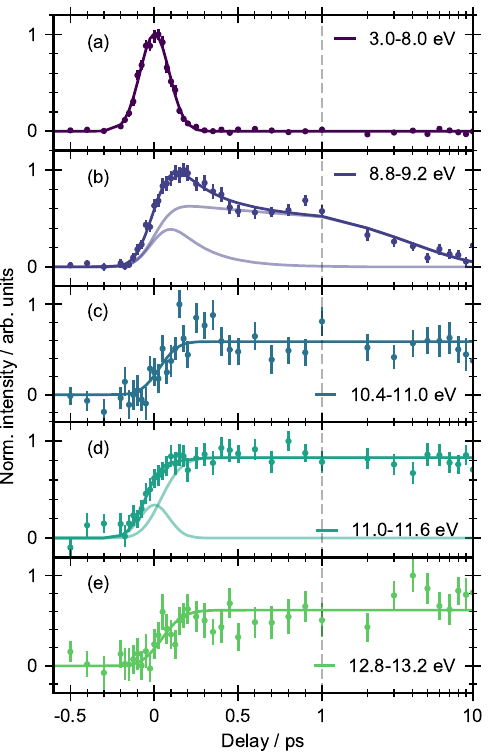}
            \caption{Integrated kinetic traces for \emph{trans}-1,2-dichloroethene with 200~nm pump and 22.3~eV probe. The error bars were obtained by standard bootstrapping analysis as described in Section \ref{Experimental}. The solid lines represent the results of fitting to models described in the text ( equations \ref{eq:par_decay} and \ref{eq:exp_rise}). For (b) and (d), the pale lines indicate the individual contributions to the multi-component fit. The dashed vertical line indicates the point from which the kinetic spectrum is plotted on a logarithmic scale.} 
            \label{fig:trans}
        \end{figure}        

        The 3D spectrum plotted in Figure \ref{fig:trans_results} allow us to identify specific regions in the photoelectron spectrum for further kinetic analysis. To determine experimental lifetimes, integrated kinetic traces were taken over the following binding energy ranges 3.0 - 8.0~eV, 8.8 - 9.2~eV, 10.4 - 11.0~eV, 11.0 - 11.6~eV and 12.8 - 13.2~eV, as shown in Figure \ref{fig:trans}. Each region was fit to a specific kinetic equation, as described below, with the results of fits given in Table \ref{tab:fits}.
        
        The energy range from 3.0 - 8.0~eV (Fig. \ref{fig:trans}a) is well fit with a single Gaussian ($\sigma$ = 87~$\pm$ 2~fs, FWHM = 205~$\pm$~5) which we take as representative of the instrument response function (IRF) of our experiment. Due to the apparent structure of the photoelectron band in this energy region fits were also made to smaller energy ranges, for example 3.0 - 6.0~eV and 6.0 - 8.0~eV, however no differences could be observed in the integrated traces or fitting results for these smaller selections.

        The energy region between 8.8 - 9.2~eV (Fig. \ref{fig:trans}b) was best fit by the sum of two exponential decays convoluted with the Gaussian IRF, which constitutes a parallel decay scheme, defined by:

        \begin{equation}
        \label{eq:par_decay}
        I(t) = (A_1e^{-\Delta{t}/\tau_1} + A_2e^{-\Delta{t}/\tau_2}) \otimes \text{IRF}
        \end{equation}
        where $A_1$ and $A_2$ represent the amplitudes of the components, $\Delta{t}$ is the delay time between pump and probe, and $\tau_1$ and $\tau_2$ are the time constants of the exponential decays. The IRF parameters were kept fixed and all other variables allowed to fit freely. The corresponding fits of the five different regions are given in Table \ref{tab:fits} and the errors corresponding to numerical uncertainties from the least-squares fitting procedure. Attempts to fit this region to a sequential kinetic model did not result in an improved fit. The time constants for the two components were found to be $\tau_1$ = 200 $\pm$ 50~fs and $\tau_2$ = 3923 $\pm$ 525~fs. We note that the signal in this region seems to not return to baseline, instead it appears to return to a constant but non-zero value. This might indicate a low-amplitude photoproduct component in this region. We chose not to integrate the 8.0 - 8.8~eV region in order to avoid overlap from the feature from the 3.0 - 8.0~eV region in the kinetic trace. 
 
        Above the first ionization energy, three long lived product signals are observed between 10.4 - 11~eV, 11.0 - 11.6~eV and 12.8 - 13.2~eV (Figure \ref{fig:trans} c - e). The integrated kinetic traces in these regions show no sign of decay within the time frame of the experiment and so their lifetimes can be considered $>$10~ps. These regions were fit with an exponential rise convoluted with the Gaussian IRF:
        \begin{equation}
        \label{eq:exp_rise}
            I(t) = A(1-e^{-\Delta{t}/\tau}) \otimes \text{IRF}
        \end{equation}
        where $\Delta{t}$ is the pump-probe time delay and $\tau$ is the rise time of the signal. An additional Gaussian with fixed parameters according to the IRF was required to fit the region 11.0 - 11.6~eV (d), indicating the presence of another short-lived state in this region. We define an upper limit of $\tau$ = 70~fs for resolvable time constants which corresponds to $\sim$ 1/3 of the IRF (FWHM). For the photoproduct regions (c) - (e), the rise time was found to be within this upper bound.

        The energy region of 10.4 - 11.0~eV shows suggestions of a peak in the trace at early time delays. The signal rises rapidly, reaching a maximum around 200~fs before dropping to a lower intensity within the first 500~fs. Therefore, we attempted to fit this region with a combination of a Gaussian and a single exponential rise convoluted with the IRF. This fit resulted in a better fit to the data by eye, but the associated errors of the fitting coefficients were larger than the coefficients themselves (Figure S9). Hence, it appears that the Gaussian component may be a reflection of low signal-to-noise in this region rather than a dynamical process of the molecule.

        The region from 12.8 - 13.2~eV shows the same behaviour, an initial rise and no appreciable decay on the time scale of the experiment. Qualitatively there appears to be additional growth in signal at longer time delays between 1 - 10~ps, however the signal-to-noise in this region is not high enough to obtain a good fit for quantitative discussion.

        In summary for \emph{trans}-1,2-dichloroethene we have observed ultrafast decay (within the experimental IRF) of the initial signal. Alongside this we see onset of signals for several products, C$_2$Cl$_2$, C$_2$HCl, C$_2$H$_2$ and Cl$^.$, which all appear rapidly, within the IRF. Finally signal is seen in the excited state region of the photoelectron spectrum which shows multiple decay processes.

        \subsection{\emph{Cis}-1,2-dichloroethene}
        
           \begin{figure*}
            \centering
            \includegraphics{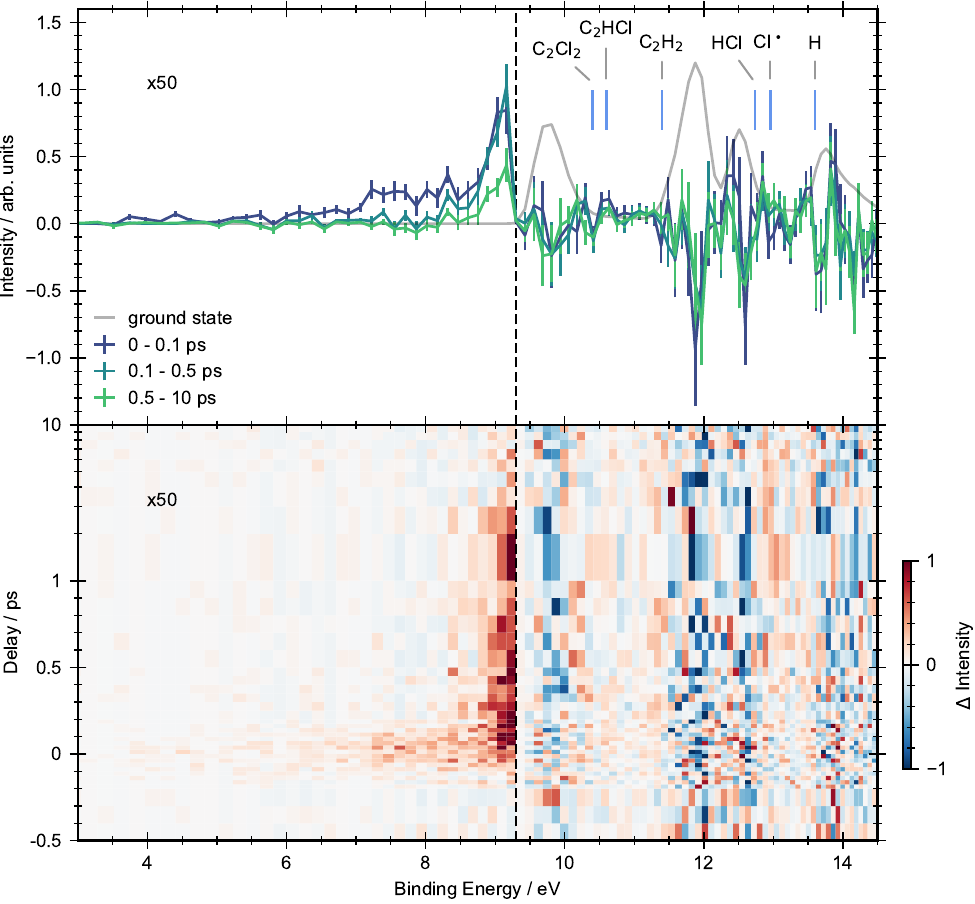}
            \caption{Pump-probe photoelectron spectra for \emph{cis}-dichloroethene with 200~nm pump and 22.3~eV probe energies.  Top panel: Grey line shows the ground-state photoelectron spectra obtained by averaging the pump-probe data at negative time delays (-1.5, -1, -0.5 and -0.4~ps). Solid lines represent the background subtracted photoelectron spectra integrated between selected pump/probe delay intervals. The error bars were obtained by standard bootstrapping analysis as described in Section \ref{Experimental}. Blue vertical lines mark the experimental vertical ionization energies of photofragments that may be produced \emph{via} photodissociation. Bottom panel: Contour plot showing the background subtracted photoelectron spectra across all pump-probe delay times. Delays greater than 1~ps are plotted on a logarithmic scale. In both panels, to aid visual interpretation the excited state signals that are observed below the onset of the first ionization band (9.3~eV, vertical dashed line) have been scaled up by a factor of $\times$50.}
             \label{fig:cis_results}
        \end{figure*}
        
\begin{figure}[h]
            \centering
            \includegraphics{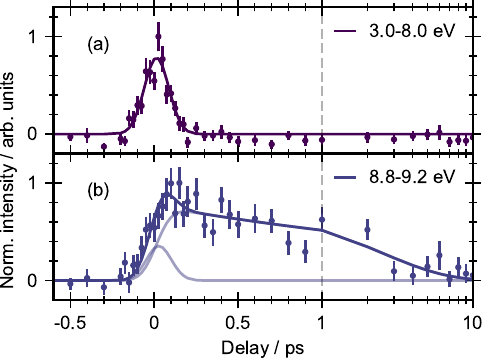}
            \caption{Integrated delay curves for \emph{cis}-1,2-dichloroethene with 200~nm pump and 21~eV probe. The dashed line indicates the point from which the kinetic spectrum is plotted on a logarithmic scale. The pale lines indicate the individual contributions to the multi-component fit}
            \label{fig:cis}
        \end{figure}

        A heat map of the time-dependent photoelectron spectra for \emph{cis}-1,2-dichloroethene obtained following excitation with a 200~nm pump along with time-averaged photoelectron spectra is shown in \ref{fig:cis_results}. The \emph{cis} isomer has a lower photoabsorption cross-section at the photon energy used compared to the \emph{trans} isomer so has a lower contrast between pump and probe signal. However, the overall excited state photoelectron spectrum still shows many similarities to the \emph{trans} results. In particular the excited state region below the first IE shows a good signal-to-noise ratio and has similar kinetic behaviour to \emph{trans}-1,2-dichloroethene. In the spectrum there is an energetically broad but short-lived feature at low binding energy (3.0 - 8.0~eV) which decays within the IRF, and a feature at higher binding energy (8.8 - 9.2~eV) which decays on a longer timescale. The integrated kinetic traces for these two regions are shown in Figure \ref{fig:cis}. They were fit according to the same procedure as described for the \emph{trans} isomer using equation \ref{eq:par_decay}, and the fitting results are also given in Table \ref{tab:fits}. The Gaussian width for the 3.0 - 8.0 eV region for the \emph{cis} dataset is slightly shorter (73~fs) than for the \emph{trans} isomer (87~fs). In the 8.8 - 9.2~eV region, both time constants are faster for the \emph{cis} isomer. $\tau_1$ is well within the IRF, and $\tau_2$ = 2.5~ps. We note that that the $\tau_1$ constant is much faster for the \emph{cis} isomer versus the \emph{trans}.
     
        Product signals that were observed for the \emph{trans} isomer are much weaker or absent for the \emph{cis} measurements. The signal-to-noise in the product regions was too low to allow fitting of a kinetic model, therefore, we do not report any fits to the product channels here. Plots of the integrated traces for the energy regions where the product channels should occur are given in the SI in Figure S6.

        \begin{table*}[h]
        \centering
            \caption{Results of fitting integrated decay curves to equations\ref{eq:par_decay}  and \ref{eq:exp_rise} for regions below the ionization potential for \emph{trans}- and \emph{cis}-dichloroethene.}

            \begin{tabular}{lllll}
                \hline
                    eBE / eV & function & \emph{trans} & \emph{cis} & Assignment \\ 
                            & & parameter / fs & parameter / fs &\\ \hline
                    3.0 - 8.0 & Gaussian& $\sigma$ = 87 $\pm$ 2 & $\sigma$ = 73 $\pm$ 5 & ES\\
                    8.8 - 9.2 & eq \ref{eq:par_decay}& $\tau_1$ = 200 $\pm$ 50 & $\tau_1$ $\leq$ 70 &\\ 
                    & & $\tau_2$ = 3923 $\pm$ 525 & $\tau_2$ = 2506 $\pm$ 556 &\\ 
                    10.4 - 11.0 & eq \ref{eq:exp_rise}& $\tau$ $\leq$ 70 & &C\textsubscript{2}HCl \\ 
                    11.0 - 11.6 & eq \ref{eq:exp_rise}& $\tau$ $\leq$ 70 & &C\textsubscript{2}H\textsubscript{2} \\ 
                    12.8 - 13.2 & eq \ref{eq:exp_rise}& $\tau$ $\leq$ 70 & &Cl$^.$ \\
                    \hline

            \end{tabular}
            \label{tab:fits}
        \end{table*}
    
    \section{Discussion}

        To aid the understanding of the time-resolved measurements it is first useful to consider what is known about the photochemistry of the 1,2-dichloroethenes. SAC-CI calculations and TD-DFT calculations show that the the main peak in the VUV absorption spectrum of the 1,2-dichloroethenes (6.27~eV for \emph{trans}, 6.62~eV for \emph{cis}) is due to a $\pi\rightarrow\pi^*$ transition from the HOMO.\cite{Arul_2008,Khvostenko2014,Locht2019-cis,Locht2020-trans} In both isomers this intense peak is found to have an underlying structure which Locht assigns as part of a Rydberg series converging to the first ionization limit mixed with n$_{Cl}$ and $\sigma^*_{C-H}$ character.\cite{Locht2019-cis,Locht2020-trans}  In the \emph{cis} isomer there is also a very weak absorption signal between 5 - 6~eV, before the rise in absorption due to the $\pi\rightarrow\pi^*$ transition. This weak signal is assigned to a B$_1$ transition which again has a mixed Rydberg, n$_{Cl}$ and $\sigma^*_{C-H}$ character.\cite{Locht2019-cis} No such transition is observed experimentally for \emph{trans}-1,2-dichloroethene, though calculations suggest that the state is present but with zero oscillator strength, and so could be involved in post-absorption dynamics. Our EOM-CCSD calculations are summarised in Tables \ref{tab_trans_eng} and \ref{tab_cis_eng}. Our values are in qualitative agreement with the previously published calculations.\cite{Locht2019-cis,Locht2020-trans} There are some differences in the energies and the precise mixed characters of the Rydberg states. These differences in character probably arise due to the use of different calculation methods and basis sets. From this supporting theory we therefore expect that initial excitation will populate the $\pi^*$ state for both isomers. However, as the \emph{cis} isomer is populated  about 0.4~eV below the maximum of the absorption peak it will have a lower amount of vibrational energy following excitation than the \emph{trans} isomer.
        
         \begin{figure}
            \centering
            \includegraphics{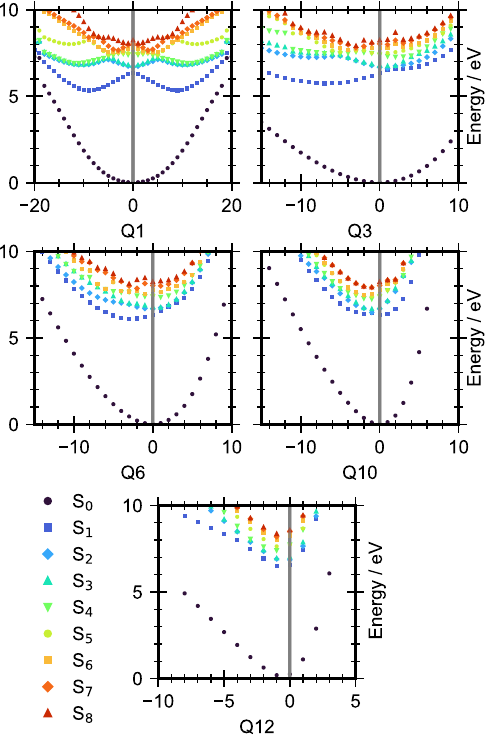}
            \caption{Cuts through the potential energy surface of \emph{trans}-1,2-dichloroethene along selected mass-weighted normal modes. Calculated using EOM-CCSD/6-311G++(3df,3pd) in QChem. The number of the mode is indicated in the x-axis label.}
            \label{fig:key_cuts}
        \end{figure}
        
        Due to the higher experimental signal-to-noise ratio in its data we will focus on the \emph{trans}-1,2-dichloroethene results. The experimental results suggest that following excitation into the $\pi^*$ state population very rapidly leaves the initially populated region of this state, within 70~fs. Cuts through the potential energy surfaces along the molecule's normal modes (given in the SI, Figure S1) indicate that the following modes should have an important effect on the early time decay of population, the C-C-Cl deformation (Q$_3$), a C-Cl stretch and CH bend combination (Q$_6$), the C=C stretch (Q$_{10}$) and a C-H stretch (Q$_{12}$). Cuts of these important modes are shown in  figure \ref{fig:key_cuts}. These cuts are presented in mass weighted coordinates where a value of 0 represents the Franck-Condon point. The bright state in these plots is S$_2$. These modes show a gradient in the Franck-Condon region which will lead to rapid movement of population out of this region of the potential energy surface away from the Franck-Condon geometry. For example, for Q$_{10}$, excitation into S$_2$  gives a geometry that is not the S$_2$ minimum. Therefore, there will be motion along the Q$_{10}$ coordinate to negative values. Further, these modes all show intersections between states S$_2$ and S$_3$, in fact this pair of states are near degenerate for large portions of the PES along these modes. All the modes apart from Q$_{12}$ also show a conical intersection between S$_2$ and S$_1$ that is very close to the Franck-Condon point. This suggests that the motion excited by the gradient will lead to rapid population transfer between the lowest three excited states (S$_1$, S$_2$ and S$_3$). Therefore, we assign the excited state feature at 3.0 - 8.0~eV, that shows very rapid decay, to transfer of the initially formed population from out of the Franck-Condon region of the S$_2$ state with rapid internal conversion to S$_1$ and S$_3$.  Following this initial rapid motion we expect that motion along the torsional mode (Q$_1$) will become important for transferring population from the excited states onto the ground state. As is clear in Figure \ref{fig:key_cuts}, at large rotations around the double bond (Q$_1$) the ground state becomes degenerate with the excited states. We would expect photoelectron signal associated with the `hot' ground state products formed from this rotation to have more excess vibrational energy than the initial ground state molecules. Therefore we expect this 'hot' signal to be observed with binding energies close to but less than the rising edge of the first peak in the ground state PES. The importance of these modes is also shown as they are known to be key in the photochemical processes of ethene and 1,1-difluoroethene.\cite{Gomez2023_difluoro,Gomez2024_ethene} where pyramidalisation of the C-H bond is coupled with twisting around the double bond to couple the excited electronic states to the ground state.

       The next feature of the time-resolved photoelectron spectrum that needs assigning is between 8.8 - 9.2~eV. This feature forms rapidly (within the IRF) and decays with two time constants of 200~fs and 4~ps. Both of these decays are associated with the peak becoming more narrow in energy (see Figure \ref{fig:trans_results}). There is also a hint that this feature does not return to baseline but instead there is a weak but constant signal left at longer times. This feature could arise from an excited state or from vibrationally excited ground-state, given its position in energy just below the onset of direct ground-state ionization. We therefore tentatively assign this feature to hot ground state, which then decays \emph{via} two pathways. This assignment is supported by a recent TRPES study on ethene with a 160~nm pump to excite into the S$_2$ state, and XUV probe (21.7~eV) to monitor the ensuing dynamics.\cite{Karashima2022} The shape of the photoelectron bands and the integrated decay curves in the energy region below onset of ground-state ionization are strikingly similar to the observations in the present study, indicating that the presence of substituents on the C=C double bond may have little effect on the initial excited state dynamics. Karashima \emph{et al.} determined that weak lower binding energy signals (3 - 6~eV) are due to the initially excited state decaying to a conical intersection on the order of 10~fs. The photoelectron signal for this conical intersection then gives rise to a more intense peak at higher binding energy (6 - 7.5~eV), which subsequently decays within 30~fs. A higher energy peak (8.5 - 10~eV) is observed which decays with two distinct time constants (0.87~ps and > 5~ps) and was attributed to decay of vibrationally excited ground state products.  It is therefore plausible that the feature in the 8.8 - 9.2~eV region in the present study of DCE is hot ground state formed within the IRF. There is currently some controversy over the theoretical assignments of these peaks in ethene.\cite{Gomez2024_ethene,Martinez_ethene} More work is needed to understand what each of these features are.

       We now turn to the photofragment signals at higher binding energy to elucidate more information on the dissociation mechanism. For the 11.0 - 11.6~eV region, the integrated trace is well fit with the sum of a Gaussian function (with parameters equal to the IRF) and an exponential rise. This Gaussian component implies the presence of a very short lived state that forms rapidly and decays within the IRF. We ascribe this to a higher lying cation state of the initial excited state signal observed in the 3.0 - 8.0~eV region. The exponential rise is associated with the formation of acetylene (\ce{C2H2}) since the ionization energy for this fragment has been measured as 11.40~eV.\cite{nistwebn} Acetylene was observed to form rapidly with respect to the initial excitation (\textless~70~fs).  Previous studies have suggested that acetylene is formed \emph{via} the secondary dissociation of the chlorovinyl radical (C\textsubscript{2}H\textsubscript{2}Cl$^.$) which itself is produced by the initial C-Cl scission from the parent molecule.\cite{Sato1997,Seki2011} Using RRK theory, Sato \emph{et al.} evaluated the rate constant for the C\textsubscript{2}H\textsubscript{2}Cl$^.$ $\rightarrow$ C\textsubscript{2}H\textsubscript{2} + Cl reaction to be 100~fs.\cite{Sato1997} The immediate presence of signal corresponding to acetylene in this study is in line with this calculation and suggests that if this sequential process occurs, it is completed within the time-resolution of the experiment. 
        
        The positive feature between 12.8 - 13.2~eV is attributed to Cl atom formation, given its measured ionization value of 12.96~eV.\cite{nistwebn} The integrated trace in this region shows a rapid growth of signal. Previous measurements have measured the kinetic energy and angular distributions of the chlorine atoms produced following photolysis at $\sim$200~nm and found two components.\cite{Umemoto1985_Cl_dis,Mo1992,Sato1993,Suzuki1994_ionimage,Huang1995,Hua2010} One pathway is from a direct crossing between the initially prepared $\pi\pi^*$ state and nearby $n\sigma^*$ and $\pi\sigma^*$ states, giving rise to a Gaussian kinetic energy distribution with angular anisotropy. The other is from dissociation of hot ground state following internal conversion and hence has a more statistical Boltzmann distribution with weak anisotropy. We would therefore expect to see two components in the kinetics of the Cl atom, an initial fast rise followed by a slower growth. We note that the Cl signal qualitatively shows what could be a rise in signal at longer times. If this is a real rise and not an artefact of low signals these would indicate two formation time constants for the chlorine atom, in line with previous measurements. This would be a fast formation step, with the chlorine atom formed by rapid decay from an excited state, possibly \emph{via} the ground state. The slower decay which appears to grow in around 3 - 4~ps would then be decay from the hot ground state. The fit with an additional time-constant is shown in Figure S8 however the signal-to-noise ratio in this region is not high enough to make any definitive assignments. 

        The energy range between 10.4 - 11.0~eV is primarily associated with C$_2$HCl with a contribution at low energy from C$_2$Cl$_2$. The signal in this region forms within the IRF. Previous studies suggest that C$_2$HCl is produced \emph{via} molecular elimination from the hot ground state. HCl, its dissociative partner, is expected to be observed at 12.74~eV however is not clearly detected due to overlap with the third peak of the ground-state photoelectron spectrum.
    
        Comparing the results for the two isomers we note that they have similar overall dynamics following photoexcitation. There is an ultrafast decay (within the experimental IRF) in the 3.0 - 8.0~eV region, while two decays are observed in the 8.8 - 9.2~eV region. The two components in this region are faster for the \emph{cis} isomer than for the \emph{trans}. In the case of $\tau_1$, this is considerably faster. This is an indication that the population transfers these decays represent must be encouraged by having the two Cl atoms (or equivalently two H atoms) close to each other. Given the speed of the decays it is motion of the H atoms that is most likely to be important, though the motion of the Cl atoms will be important for the longer timescales. Further, these differences in decay rates could be projected onto differences in the ratios of product channels between the isomers. For example, Sato \cite{Sato1997} saw that the relative yields for HCl from \emph{cis} and \emph{trans} was 0.3 and 0.5 respectively. As we do not observe clear product signals for \emph{cis}-1,2-dichloroethene it is not possible for us to compare relative signals between the isomers.

        That the time constant for the initial decay from the excited state is very similar shows that the early-time excited state dynamics are not affected by the relative position of the substituents around the double bond. This suggests that these early dynamics are controlled by the motions of the C and H atoms and not those of the Cl atoms which seem to be spectators. This is further shown by the similarity between the ethene and chloroethene dynamics.\cite{Karashima2022} Therefore, we believe the initial dynamics in the 1,2-dichloroethenes is the same as in ethene. Rapid decay from the Franck-Condon point to a conical intersection with the ground state. The population transfer is then encouraged by lengthening of the C=C bond and out-of-plane motion of the H atoms.\cite{Martinez_ethene,Karashima2022} The early-time dynamics are also not affected by the swapping of character of S$_2$ and S$_3$ between the two isomers. Showing that the second Rydberg state, with mixed n$_{Cl}$ and Rydberg character has little effect on the photodynamics. This is inline with studies on ethene dynamics where it was found that the low-lying ethene Rydberg state had little effect on the early time dynamics.\cite{Martinez_ethene}  

        In summary, we believe the following dynamics occur following photoexcitation of 1,2-dichloroethene, based upon previous studies of ethene.\cite{Martinez_ethene, Karashima2022} The initial photon absorption is a $\pi\rightarrow\pi^*$ transition into the S$_2$ (S$_3$ for \emph{cis}) state. The population from this subsequently decays on an ultrafast time scale to a region of the potential energy surface which has strong coupling to the ground state. This population transfer is facilitated by initial stretching of the C=C bond and pyramidalization around one of the carbon atoms bonds, largely driven by the motion of the H atoms. We believe, that given that all products appear to be formed at the same time and that it is known that HCl is formed on the ground state, that it is from the ground-state that the detected photoproducts are formed.
        
        One aim of this study was to look for signs of isomerisation between \emph{trans} and \emph{cis}-1,2-dichloroethene. Though no conclusive evidence is found there are hints in the data. One hint of isomerisation occurring includes the second peak in the ground-state photoelectron spectrum showing a rise in signal at a lower binding energy in the \emph{cis} data than in the \emph{trans} data. In this energy region there is clear signal due to formation of C$_2$H$_2$, but there is possibly a second feature just before the edge of the ground-state signal (see Figure \ref{fig:trans_results}). There are also hints in the ground-state bleaches of differences between the the first and second ground-state photoelectron peaks that could be due to the change in intensity of these peaks when going from \emph{trans} to \emph{cis}-1,2-dichloroethene. Further studies are required to look for signs of such isomerisation.

        The other aim of this study was to improve our understanding of the dynamics that occur following photoexcitation of the 1,2-dichloroethenes. Our results show, that at least for \emph{trans}-1,2-dichloroethene, the photoproducts are all formed very rapidly, within the time resolution of the experiments. As the formation is so rapid and it is known that HCl is formed on the ground state we suggest that all products are formed following ultrafast population of the ground-state. This is in contrast to previous static measurements that assumed that C$_2$H$_2$ was formed \emph{via} a curve crossing on the excited state. To be sure of the assigned formation mechanisms of both C$_2$H$_2$ and C$_2$HCl more work is needed. Experimentally, measurements with higher electron energy resolution where the photoelectrons and photoions are measured in coincidence would be useful. Such measurements would allow clear and unambiguous assignment of each peak observed in the photoelectron spectrum to a particular mass. On the theoretical side, \emph{ab inito} dynamic calculations would help to reveal the mechanisms in more depth.  

    \section{Conclusions}        
        In conclusion we have studied the photochemistry of the two isomers of 1,2-dichloroethene using time resolved photoelectron spectroscopy coupled with theoretical calculations of the potential energy surface. Following excitation into the $\pi^*$ state there is an ultrafast decay from the Franck-Condon region. This decay is caused by gradients along modes for C=C, C-Cl and C-H stretches and the C-C-H bend. Following this movement out of the Franck-Condon region there is rapid population transfer onto the ground state. From this ground state the formation of several products are observed. Signatures are detected for formation of C$_2$H$_2$, C$_2$Cl$_2$ and Cl$^.$. The formation of all these products is observed to occur on an ultrafast timescale.

\section*{Author Contributions}
        The experiment was conceived by MAP with input on the planning and implementation from RAI and RSM. MAP, RAI, RSM, DAH, HGM, HJT collected the data with support from JT, YZ,  ASW, RTC and ES who run and manage the Artemis laser facility. The experimental data was analysed by HGM and MAP, and the theoretical calculations performed by MAP. The combined output from the experiment and calculations were interpreted by MAP, RAI, RSM, DAH, HGM. The manuscript was written by MAP and HGM with input and edits from all authors.

\section*{Conflicts of interest}
    There are no conflicts to declare.

\section*{Acknowledgements}
    The authors thank the STFC for access to the Artemis facility. RSM thanks the EPSRC for financial support (EP/X027635/1). DAH thanks The Netherlands Organization for Scientific Research (NWO) for support under grant numbers STU.019.009 and VI-VIDI-193.037. HJT is grateful to the STFC XFEL Hub for Physical Sciences and the University of Southampton for a studentship. RAI and HMG acknowledge EPSRC for a studentship (EP/N509577/1 and EP/T517793/1).



\balance


\bibliography{Article} 

\providecommand*{\mcitethebibliography}{\thebibliography}
\csname @ifundefined\endcsname{endmcitethebibliography}
{\let\endmcitethebibliography\endthebibliography}{}
\begin{mcitethebibliography}{35}
\providecommand*{\natexlab}[1]{#1}
\providecommand*{\mciteSetBstSublistMode}[1]{}
\providecommand*{\mciteSetBstMaxWidthForm}[2]{}
\providecommand*{\mciteBstWouldAddEndPuncttrue}
  {\def\EndOfBibitem{\unskip.}}
\providecommand*{\mciteBstWouldAddEndPunctfalse}
  {\let\EndOfBibitem\relax}
\providecommand*{\mciteSetBstMidEndSepPunct}[3]{}
\providecommand*{\mciteSetBstSublistLabelBeginEnd}[3]{}
\providecommand*{\EndOfBibitem}{}
\mciteSetBstSublistMode{f}
\mciteSetBstMaxWidthForm{subitem}
{(\emph{\alph{mcitesubitemcount}})}
\mciteSetBstSublistLabelBeginEnd{\mcitemaxwidthsubitemform\space}
{\relax}{\relax}

\bibitem[Erez \emph{et~al.}(2014)Erez, Presiado, Gepshtein, Simkovitch, and Huppert]{retinalTR}
Y.~Erez, I.~Presiado, R.~Gepshtein, R.~Simkovitch and D.~Huppert, \emph{J. Mod. Opt.}, 2014, \textbf{61}, 1589--1604\relax
\mciteBstWouldAddEndPuncttrue
\mciteSetBstMidEndSepPunct{\mcitedefaultmidpunct}
{\mcitedefaultendpunct}{\mcitedefaultseppunct}\relax
\EndOfBibitem
\bibitem[Karashima \emph{et~al.}(2022)Karashima, Humeniuk, Glover, and Suzuki]{Karashima2022}
S.~Karashima, A.~Humeniuk, W.~J. Glover and T.~Suzuki, \emph{J. Phys. Chem. A}, 2022, \textbf{126}, 3873--3879\relax
\mciteBstWouldAddEndPuncttrue
\mciteSetBstMidEndSepPunct{\mcitedefaultmidpunct}
{\mcitedefaultendpunct}{\mcitedefaultseppunct}\relax
\EndOfBibitem
\bibitem[Wang \emph{et~al.}(2022)Wang, Waters, Zhang, Suchan, Svoboda, Luu, Perry, Yin, Slav{\'\i}{\v{c}}ek, and W{\"o}rner]{wang2022stilbene}
C.~Wang, M.~D. Waters, P.~Zhang, J.~Suchan, V.~Svoboda, T.~T. Luu, C.~Perry, Z.~Yin, P.~Slav{\'\i}{\v{c}}ek and H.~J. W{\"o}rner, \emph{Nat. Chem.}, 2022, \textbf{14}, 1126--1132\relax
\mciteBstWouldAddEndPuncttrue
\mciteSetBstMidEndSepPunct{\mcitedefaultmidpunct}
{\mcitedefaultendpunct}{\mcitedefaultseppunct}\relax
\EndOfBibitem
\bibitem[Sulbaek~Andersen \emph{et~al.}(2022)Sulbaek~Andersen, Volkova, Hass, Lengkong, Hovanessian, Sølling, Wallington, and Nielsen]{atmos_chloros}
M.~P. Sulbaek~Andersen, A.~Volkova, S.~A. Hass, J.~W. Lengkong, D.~Hovanessian, T.~I. Sølling, T.~J. Wallington and O.~J. Nielsen, \emph{Phys. Chem. Chem. Phys.}, 2022, \textbf{24}, 7356--7373\relax
\mciteBstWouldAddEndPuncttrue
\mciteSetBstMidEndSepPunct{\mcitedefaultmidpunct}
{\mcitedefaultendpunct}{\mcitedefaultseppunct}\relax
\EndOfBibitem
\bibitem[Yamada \emph{et~al.}(2001)Yamada, El-Sinawi, Siraj, Taylor, Peng, Hu, and Marshall]{OH-Yamada2001}
T.~Yamada, A.~El-Sinawi, M.~Siraj, P.~H. Taylor, J.~Peng, X.~Hu and P.~Marshall, \emph{J. Phys. Chem. A}, 2001, \textbf{105}, 7588--7597\relax
\mciteBstWouldAddEndPuncttrue
\mciteSetBstMidEndSepPunct{\mcitedefaultmidpunct}
{\mcitedefaultendpunct}{\mcitedefaultseppunct}\relax
\EndOfBibitem
\bibitem[Canosa-Mas \emph{et~al.}(2001)Canosa-Mas, Dillon, Sidebottom, Thompson, and Wayne]{OH-CanosaMas2001}
C.~E. Canosa-Mas, T.~J. Dillon, H.~Sidebottom, K.~C. Thompson and R.~P. Wayne, \emph{Phys. Chem. Chem. Phys.}, 2001, \textbf{3}, 542--550\relax
\mciteBstWouldAddEndPuncttrue
\mciteSetBstMidEndSepPunct{\mcitedefaultmidpunct}
{\mcitedefaultendpunct}{\mcitedefaultseppunct}\relax
\EndOfBibitem
\bibitem[Locht \emph{et~al.}(2019)Locht, Dehareng, and Leyh]{Locht2019-cis}
R.~Locht, D.~Dehareng and B.~Leyh, \emph{AIP Adv.}, 2019, \textbf{9}, 015305\relax
\mciteBstWouldAddEndPuncttrue
\mciteSetBstMidEndSepPunct{\mcitedefaultmidpunct}
{\mcitedefaultendpunct}{\mcitedefaultseppunct}\relax
\EndOfBibitem
\bibitem[Locht \emph{et~al.}(2020)Locht, Dehareng, and Leyh]{Locht2020-trans}
R.~Locht, D.~Dehareng and B.~Leyh, \emph{J. Quant. Spec. Rad. Trans.}, 2020, \textbf{251}, 107048\relax
\mciteBstWouldAddEndPuncttrue
\mciteSetBstMidEndSepPunct{\mcitedefaultmidpunct}
{\mcitedefaultendpunct}{\mcitedefaultseppunct}\relax
\EndOfBibitem
\bibitem[Wijnen(1961)]{Wijnen1961}
M.~J. Wijnen, \emph{J. Am. Chem. Soc.}, 1961, \textbf{83}, 4109--4110\relax
\mciteBstWouldAddEndPuncttrue
\mciteSetBstMidEndSepPunct{\mcitedefaultmidpunct}
{\mcitedefaultendpunct}{\mcitedefaultseppunct}\relax
\EndOfBibitem
\bibitem[Ausubel and Wijnen(1975)]{Ausubel1975trans}
R.~Ausubel and M.~Wijnen, \emph{J. Photochem.}, 1975, \textbf{4}, 241--248\relax
\mciteBstWouldAddEndPuncttrue
\mciteSetBstMidEndSepPunct{\mcitedefaultmidpunct}
{\mcitedefaultendpunct}{\mcitedefaultseppunct}\relax
\EndOfBibitem
\bibitem[Ausubel and Wijnen(1975)]{Ausubel1975cis}
R.~Ausubel and M.~H.~J. Wijnen, \emph{Int. J. Chem. Kin.}, 1975, \textbf{7}, 739--751\relax
\mciteBstWouldAddEndPuncttrue
\mciteSetBstMidEndSepPunct{\mcitedefaultmidpunct}
{\mcitedefaultendpunct}{\mcitedefaultseppunct}\relax
\EndOfBibitem
\bibitem[Berry(1974)]{Berry1974}
M.~J. Berry, \emph{J. Chem. Phys.}, 1974, \textbf{61}, 3114--3143\relax
\mciteBstWouldAddEndPuncttrue
\mciteSetBstMidEndSepPunct{\mcitedefaultmidpunct}
{\mcitedefaultendpunct}{\mcitedefaultseppunct}\relax
\EndOfBibitem
\bibitem[Moss \emph{et~al.}(1981)Moss, Ensminger, and McDonald]{Moss1981}
M.~G. Moss, M.~D. Ensminger and J.~D. McDonald, \emph{J. Chem. Phys.}, 1981, \textbf{74}, 6631--6635\relax
\mciteBstWouldAddEndPuncttrue
\mciteSetBstMidEndSepPunct{\mcitedefaultmidpunct}
{\mcitedefaultendpunct}{\mcitedefaultseppunct}\relax
\EndOfBibitem
\bibitem[Umemoto \emph{et~al.}(1985)Umemoto, Seki, Shinohara, Nagashima, Nishi, Kinoshita, and Shimada]{Umemoto1985_Cl_dis}
M.~Umemoto, K.~Seki, H.~Shinohara, U.~Nagashima, N.~Nishi, M.~Kinoshita and R.~Shimada, \emph{J. Chem. Phys.}, 1985, \textbf{83}, 1657--1666\relax
\mciteBstWouldAddEndPuncttrue
\mciteSetBstMidEndSepPunct{\mcitedefaultmidpunct}
{\mcitedefaultendpunct}{\mcitedefaultseppunct}\relax
\EndOfBibitem
\bibitem[Mo \emph{et~al.}(1992)Mo, Tonokura, Matsumi, Kawasaki, Sato, Arikawa, Reilly, Xie, Yang, Huang, and Gordon]{Mo1992}
Y.~Mo, K.~Tonokura, Y.~Matsumi, M.~Kawasaki, T.~Sato, T.~Arikawa, P.~T.~A. Reilly, Y.~Xie, Y.-a. Yang, Y.~Huang and R.~J. Gordon, \emph{J. Chem. Phys.}, 1992, \textbf{97}, 4815--4826\relax
\mciteBstWouldAddEndPuncttrue
\mciteSetBstMidEndSepPunct{\mcitedefaultmidpunct}
{\mcitedefaultendpunct}{\mcitedefaultseppunct}\relax
\EndOfBibitem
\bibitem[Sato \emph{et~al.}(1993)Sato, Shihira, Tsunashima, Umemoto, Takayanagi, Furukawa, and Ohno]{Sato1993}
K.~Sato, Y.~Shihira, S.~Tsunashima, H.~Umemoto, T.~Takayanagi, K.~Furukawa and S.~I. Ohno, \emph{J. Chem.Phys.}, 1993, \textbf{99}, 1703--1709\relax
\mciteBstWouldAddEndPuncttrue
\mciteSetBstMidEndSepPunct{\mcitedefaultmidpunct}
{\mcitedefaultendpunct}{\mcitedefaultseppunct}\relax
\EndOfBibitem
\bibitem[Suzuki \emph{et~al.}(1994)Suzuki, Tonokura, Bontuyan, and Hashimoto]{Suzuki1994_ionimage}
T.~Suzuki, K.~Tonokura, L.~S. Bontuyan and N.~Hashimoto, \emph{J. Phys. Chem.}, 1994, \textbf{98}, 13447--13451\relax
\mciteBstWouldAddEndPuncttrue
\mciteSetBstMidEndSepPunct{\mcitedefaultmidpunct}
{\mcitedefaultendpunct}{\mcitedefaultseppunct}\relax
\EndOfBibitem
\bibitem[Huang \emph{et~al.}(1995)Huang, Yang, He, Hashimoto, and Gordon]{Huang1995}
Y.~Huang, Y.~A. Yang, G.~He, S.~Hashimoto and R.~J. Gordon, \emph{J. Chem. Phys.}, 1995, \textbf{103}, 5476--5487\relax
\mciteBstWouldAddEndPuncttrue
\mciteSetBstMidEndSepPunct{\mcitedefaultmidpunct}
{\mcitedefaultendpunct}{\mcitedefaultseppunct}\relax
\EndOfBibitem
\bibitem[Hua \emph{et~al.}(2010)Hua, Zhang, Lee, Chao, Zhang, and Lin]{Hua2010}
L.~Hua, X.~Zhang, W.~B. Lee, M.~H. Chao, B.~Zhang and K.~C. Lin, \emph{J.f Phys. Chem. A}, 2010, \textbf{114}, 37--44\relax
\mciteBstWouldAddEndPuncttrue
\mciteSetBstMidEndSepPunct{\mcitedefaultmidpunct}
{\mcitedefaultendpunct}{\mcitedefaultseppunct}\relax
\EndOfBibitem
\bibitem[He \emph{et~al.}(1993)He, Yang, Huang, and Gordon]{He1993}
G.~X. He, Y.~A. Yang, Y.~Huang and R.~J. Gordon, \emph{J. Phys. Chem.}, 1993, \textbf{97}, 2186--2193\relax
\mciteBstWouldAddEndPuncttrue
\mciteSetBstMidEndSepPunct{\mcitedefaultmidpunct}
{\mcitedefaultendpunct}{\mcitedefaultseppunct}\relax
\EndOfBibitem
\bibitem[Sato \emph{et~al.}(1997)Sato, Tsunashima, Takayanagi, Fujisawa, and Yokoyama]{Sato1997}
K.~Sato, S.~Tsunashima, T.~Takayanagi, G.~Fujisawa and A.~Yokoyama, \emph{J. Chem. Phys}, 1997, \textbf{106}, 10123--10133\relax
\mciteBstWouldAddEndPuncttrue
\mciteSetBstMidEndSepPunct{\mcitedefaultmidpunct}
{\mcitedefaultendpunct}{\mcitedefaultseppunct}\relax
\EndOfBibitem
\bibitem[Seki \emph{et~al.}(2011)Seki, Kobayashi, and Ebata]{Seki2011}
K.~Seki, T.~Kobayashi and K.~Ebata, \emph{J. Photochem. Photobio. A}, 2011, \textbf{219}, 200--203\relax
\mciteBstWouldAddEndPuncttrue
\mciteSetBstMidEndSepPunct{\mcitedefaultmidpunct}
{\mcitedefaultendpunct}{\mcitedefaultseppunct}\relax
\EndOfBibitem
\bibitem[Stolow \emph{et~al.}(2004)Stolow, Bragg, and Neumark]{Stolow2004}
A.~Stolow, A.~E. Bragg and D.~M. Neumark, \emph{Chem. Revs.}, 2004, \textbf{104}, 1719--1758\relax
\mciteBstWouldAddEndPuncttrue
\mciteSetBstMidEndSepPunct{\mcitedefaultmidpunct}
{\mcitedefaultendpunct}{\mcitedefaultseppunct}\relax
\EndOfBibitem
\bibitem[Smith \emph{et~al.}(2018)Smith, Warne, Bellshaw, Horke, Tudorovskya, Springate, Jones, Cacho, Chapman, Kirrander, and Minns]{Smith:PRL120:183003}
A.~D. Smith, E.~M. Warne, D.~Bellshaw, D.~A. Horke, M.~Tudorovskya, E.~Springate, A.~J.~H. Jones, C.~Cacho, R.~T. Chapman, A.~Kirrander and R.~S. Minns, \emph{Phys. Rev. Lett.}, 2018, \textbf{120}, 183003\relax
\mciteBstWouldAddEndPuncttrue
\mciteSetBstMidEndSepPunct{\mcitedefaultmidpunct}
{\mcitedefaultendpunct}{\mcitedefaultseppunct}\relax
\EndOfBibitem
\bibitem[Pathak \emph{et~al.}(2020)Pathak, Ibele, Boll, Callegari, Demidovich, Erk, Feifel, Forbes, {Di Fraia}, Giannessi, Hansen, Holland, Ingle, Mason, Plekan, Prince, Rouz{\'{e}}e, Squibb, Tross, Ashfold, Curchod, and Rolles]{Pathak2020}
S.~Pathak, L.~M. Ibele, R.~Boll, C.~Callegari, A.~Demidovich, B.~Erk, R.~Feifel, R.~Forbes, M.~{Di Fraia}, L.~Giannessi, C.~S. Hansen, D.~M. Holland, R.~A. Ingle, R.~Mason, O.~Plekan, K.~C. Prince, A.~Rouz{\'{e}}e, R.~J. Squibb, J.~Tross, M.~N. Ashfold, B.~F. Curchod and D.~Rolles, \emph{Nat. Chem.}, 2020, \textbf{12}, 795--800\relax
\mciteBstWouldAddEndPuncttrue
\mciteSetBstMidEndSepPunct{\mcitedefaultmidpunct}
{\mcitedefaultendpunct}{\mcitedefaultseppunct}\relax
\EndOfBibitem
\bibitem[Borne \emph{et~al.}(2024)Borne, Cooper, Ashfold, Bachmann, Bhattacharyya, Boll, Bonanomi, Bosch, Callegari, Centurion, Coreno, Curchod, Danailov, Demidovich, {Di Fraia}, Erk, Faccial{\`{a}}, Feifel, Forbes, Hansen, Holland, Ingle, Lindh, Ma, McGhee, Muvva, Nunes, Odate, Pathak, Plekan, Prince, Rebernik, Rouz{\'{e}}e, Rudenko, Simoncig, Squibb, Venkatachalam, Vozzi, Weber, Kirrander, and Rolles]{Borne2024}
K.~D. Borne, J.~C. Cooper, M.~N. Ashfold, J.~Bachmann, S.~Bhattacharyya, R.~Boll, M.~Bonanomi, M.~Bosch, C.~Callegari, M.~Centurion, M.~Coreno, B.~F. Curchod, M.~B. Danailov, A.~Demidovich, M.~{Di Fraia}, B.~Erk, D.~Faccial{\`{a}}, R.~Feifel, R.~J. Forbes, C.~S. Hansen, D.~M. Holland, R.~A. Ingle, R.~Lindh, L.~Ma, H.~G. McGhee, S.~B. Muvva, J.~P.~F. Nunes, A.~Odate, S.~Pathak, O.~Plekan, K.~C. Prince, P.~Rebernik, A.~Rouz{\'{e}}e, A.~Rudenko, A.~Simoncig, R.~J. Squibb, A.~S. Venkatachalam, C.~Vozzi, P.~M. Weber, A.~Kirrander and D.~Rolles, \emph{Nat. Che.}, 2024, \textbf{16}, 499--505\relax
\mciteBstWouldAddEndPuncttrue
\mciteSetBstMidEndSepPunct{\mcitedefaultmidpunct}
{\mcitedefaultendpunct}{\mcitedefaultseppunct}\relax
\EndOfBibitem
\bibitem[Frassetto \emph{et~al.}(2011)Frassetto, Cacho, Froud, Turcu, Villoresi, Bryan, Springate, and Poletto]{Art_mono}
F.~Frassetto, C.~Cacho, C.~A. Froud, I.~E. Turcu, P.~Villoresi, W.~A. Bryan, E.~Springate and L.~Poletto, \emph{Opt. Express}, 2011, \textbf{19}, 19169--19181\relax
\mciteBstWouldAddEndPuncttrue
\mciteSetBstMidEndSepPunct{\mcitedefaultmidpunct}
{\mcitedefaultendpunct}{\mcitedefaultseppunct}\relax
\EndOfBibitem
\bibitem[Frisch \emph{et~al.}(2016)Frisch, Trucks, Schlegel, Scuseria, Robb, Cheeseman, Scalmani, Barone, Petersson, Nakatsuji, Li, Caricato, Marenich, Bloino, Janesko, Gomperts, Mennucci, Hratchian, Ortiz, Izmaylov, Sonnenberg, Williams-Young, Ding, Lipparini, Egidi, Goings, Peng, Petrone, Henderson, Ranasinghe, Zakrzewski, Gao, Rega, Zheng, Liang, Hada, Ehara, Toyota, Fukuda, Hasegawa, Ishida, Nakajima, Honda, Kitao, Nakai, Vreven, Throssell, Montgomery, Peralta, Ogliaro, Bearpark, Heyd, Brothers, Kudin, Staroverov, Keith, Kobayashi, Normand, Raghavachari, Rendell, Burant, Iyengar, Tomasi, Cossi, Millam, Klene, Adamo, Cammi, Ochterski, Martin, Morokuma, Farkas, Foresman, and Fox]{g16}
M.~J. Frisch, G.~W. Trucks, H.~B. Schlegel, G.~E. Scuseria, M.~A. Robb, J.~R. Cheeseman, G.~Scalmani, V.~Barone, G.~A. Petersson, H.~Nakatsuji, X.~Li, M.~Caricato, A.~V. Marenich, J.~Bloino, B.~G. Janesko, R.~Gomperts, B.~Mennucci, H.~P. Hratchian, J.~V. Ortiz, A.~F. Izmaylov, J.~L. Sonnenberg, D.~Williams-Young, F.~Ding, F.~Lipparini, F.~Egidi, J.~Goings, B.~Peng, A.~Petrone, T.~Henderson, D.~Ranasinghe, V.~G. Zakrzewski, J.~Gao, N.~Rega, G.~Zheng, W.~Liang, M.~Hada, M.~Ehara, K.~Toyota, R.~Fukuda, J.~Hasegawa, M.~Ishida, T.~Nakajima, Y.~Honda, O.~Kitao, H.~Nakai, T.~Vreven, K.~Throssell, J.~A. Montgomery, {Jr.}, J.~E. Peralta, F.~Ogliaro, M.~J. Bearpark, J.~J. Heyd, E.~N. Brothers, K.~N. Kudin, V.~N. Staroverov, T.~A. Keith, R.~Kobayashi, J.~Normand, K.~Raghavachari, A.~P. Rendell, J.~C. Burant, S.~S. Iyengar, J.~Tomasi, M.~Cossi, J.~M. Millam, M.~Klene, C.~Adamo, R.~Cammi, J.~W. Ochterski, R.~L. Martin, K.~Morokuma, O.~Farkas, J.~B. Foresman and D.~J. Fox, \emph{Gaussian˜16 {R}evision {A}.01}, 2016,
  Gaussian Inc. Wallingford CT\relax
\mciteBstWouldAddEndPuncttrue
\mciteSetBstMidEndSepPunct{\mcitedefaultmidpunct}
{\mcitedefaultendpunct}{\mcitedefaultseppunct}\relax
\EndOfBibitem
\bibitem[Shao \emph{et~al.}(2014)Shao, Gan, Epifanovsky, Gilbert, Wormit, Kussmann, Lange, Behn, Deng, Feng, Ghosh, Goldey, Horn, Jacobson, Kaliman, Khaliullin, Ku{\'{s}}, Landau, Liu, Proynov, Rhee, Richard, Rohrdanz, Steele, Sundstrom, Woodcock, Zimmerman, Zuev, Albrecht, Alguire, Austin, Beran, Bernard, Berquist, Brandhorst, Bravaya, Brown, Casanova, Chang, Chen, Chien, Closser, Crittenden, Diedenhofen, DiStasio, Do, Dutoi, Edgar, Fatehi, Fusti-Molnar, Ghysels, Golubeva-Zadorozhnaya, Gomes, Hanson-Heine, Harbach, Hauser, Hohenstein, Holden, Jagau, Ji, Kaduk, Khistyaev, Kim, Kim, King, Klunzinger, Kosenkov, Kowalczyk, Krauter, Lao, Laurent, Lawler, Levchenko, Lin, Liu, Livshits, Lochan, Luenser, Manohar, Manzer, Mao, Mardirossian, Marenich, Maurer, Mayhall, Neuscamman, Oana, Olivares-Amaya, O'Neill, Parkhill, Perrine, Peverati, Prociuk, Rehn, Rosta, Russ, Sharada, Sharma, Small, Sodt, Stein, Stück, Su, Thom, Tsuchimochi, Vanovschi, Vogt, Vydrov, Wang, Watson, Wenzel, White, Williams, Yang, Yeganeh, Yost,
  You, Zhang, Zhang, Zhao, Brooks, Chan, Chipman, Cramer, Goddard, Gordon, Hehre, Klamt, Schaefer, Schmidt, Sherrill, Truhlar, Warshel, Xu, Aspuru-Guzik, Baer, Bell, Besley, Chai, Dreuw, Dunietz, Furlani, Gwaltney, Hsu, Jung, Kong, Lambrecht, Liang, Ochsenfeld, Rassolov, Slipchenko, Subotnik, Voorhis, Herbert, Krylov, Gill, and Head-Gordon]{qChem}
Y.~Shao, Z.~Gan, E.~Epifanovsky, A.~T. Gilbert, M.~Wormit, J.~Kussmann, A.~W. Lange, A.~Behn, J.~Deng, X.~Feng, D.~Ghosh, M.~Goldey, P.~R. Horn, L.~D. Jacobson, I.~Kaliman, R.~Z. Khaliullin, T.~Ku{\'{s}}, A.~Landau, J.~Liu, E.~I. Proynov, Y.~M. Rhee, R.~M. Richard, M.~A. Rohrdanz, R.~P. Steele, E.~J. Sundstrom, H.~L. Woodcock, P.~M. Zimmerman, D.~Zuev, B.~Albrecht, E.~Alguire, B.~Austin, G.~J.~O. Beran, Y.~A. Bernard, E.~Berquist, K.~Brandhorst, K.~B. Bravaya, S.~T. Brown, D.~Casanova, C.-M. Chang, Y.~Chen, S.~H. Chien, K.~D. Closser, D.~L. Crittenden, M.~Diedenhofen, R.~A. DiStasio, H.~Do, A.~D. Dutoi, R.~G. Edgar, S.~Fatehi, L.~Fusti-Molnar, A.~Ghysels, A.~Golubeva-Zadorozhnaya, J.~Gomes, M.~W. Hanson-Heine, P.~H. Harbach, A.~W. Hauser, E.~G. Hohenstein, Z.~C. Holden, T.-C. Jagau, H.~Ji, B.~Kaduk, K.~Khistyaev, J.~Kim, J.~Kim, R.~A. King, P.~Klunzinger, D.~Kosenkov, T.~Kowalczyk, C.~M. Krauter, K.~U. Lao, A.~D. Laurent, K.~V. Lawler, S.~V. Levchenko, C.~Y. Lin, F.~Liu, E.~Livshits, R.~C. Lochan, A.~Luenser,
  P.~Manohar, S.~F. Manzer, S.-P. Mao, N.~Mardirossian, A.~V. Marenich, S.~A. Maurer, N.~J. Mayhall, E.~Neuscamman, C.~M. Oana, R.~Olivares-Amaya, D.~P. O'Neill, J.~A. Parkhill, T.~M. Perrine, R.~Peverati, A.~Prociuk, D.~R. Rehn, E.~Rosta, N.~J. Russ, S.~M. Sharada, S.~Sharma, D.~W. Small, A.~Sodt, T.~Stein, D.~Stück, Y.-C. Su, A.~J. Thom, T.~Tsuchimochi, V.~Vanovschi, L.~Vogt, O.~Vydrov, T.~Wang, M.~A. Watson, J.~Wenzel, A.~White, C.~F. Williams, J.~Yang, S.~Yeganeh, S.~R. Yost, Z.-Q. You, I.~Y. Zhang, X.~Zhang, Y.~Zhao, B.~R. Brooks, G.~K. Chan, D.~M. Chipman, C.~J. Cramer, W.~A. Goddard, M.~S. Gordon, W.~J. Hehre, A.~Klamt, H.~F. Schaefer, M.~W. Schmidt, C.~D. Sherrill, D.~G. Truhlar, A.~Warshel, X.~Xu, A.~Aspuru-Guzik, R.~Baer, A.~T. Bell, N.~A. Besley, J.-D. Chai, A.~Dreuw, B.~D. Dunietz, T.~R. Furlani, S.~R. Gwaltney, C.-P. Hsu, Y.~Jung, J.~Kong, D.~S. Lambrecht, W.~Liang, C.~Ochsenfeld, V.~A. Rassolov, L.~V. Slipchenko, J.~E. Subotnik, T.~V. Voorhis, J.~M. Herbert, A.~I. Krylov, P.~M. Gill and
  M.~Head-Gordon, \emph{Mol. Phys.}, 2014, \textbf{113}, 184--215\relax
\mciteBstWouldAddEndPuncttrue
\mciteSetBstMidEndSepPunct{\mcitedefaultmidpunct}
{\mcitedefaultendpunct}{\mcitedefaultseppunct}\relax
\EndOfBibitem
\bibitem[Arulmozhiraja \emph{et~al.}(2008)Arulmozhiraja, Ehara, and Nakatsuji]{Arul_2008}
S.~Arulmozhiraja, M.~Ehara and H.~Nakatsuji, \emph{J. Chem. Phys.}, 2008, \textbf{129}, 174506\relax
\mciteBstWouldAddEndPuncttrue
\mciteSetBstMidEndSepPunct{\mcitedefaultmidpunct}
{\mcitedefaultendpunct}{\mcitedefaultseppunct}\relax
\EndOfBibitem
\bibitem[Khvostenko(2014)]{Khvostenko2014}
O.~Khvostenko, \emph{J. Elec. Spec. Rel. Phen.}, 2014, \textbf{195}, 220--229\relax
\mciteBstWouldAddEndPuncttrue
\mciteSetBstMidEndSepPunct{\mcitedefaultmidpunct}
{\mcitedefaultendpunct}{\mcitedefaultseppunct}\relax
\EndOfBibitem
\bibitem[Gómez \emph{et~al.}(2023)Gómez, Singer, González, and Worth]{Gomez2023_difluoro}
S.~Gómez, N.~K. Singer, L.~González and G.~A. Worth, \emph{Can. J. Chem.}, 2023, \textbf{101}, 745--757\relax
\mciteBstWouldAddEndPuncttrue
\mciteSetBstMidEndSepPunct{\mcitedefaultmidpunct}
{\mcitedefaultendpunct}{\mcitedefaultseppunct}\relax
\EndOfBibitem
\bibitem[Gómez \emph{et~al.}(2024)Gómez, Spinlove, and Worth]{Gomez2024_ethene}
S.~Gómez, E.~Spinlove and G.~Worth, \emph{Phys. Chem. Chem. Phys.}, 2024, \textbf{26}, 1829--1844\relax
\mciteBstWouldAddEndPuncttrue
\mciteSetBstMidEndSepPunct{\mcitedefaultmidpunct}
{\mcitedefaultendpunct}{\mcitedefaultseppunct}\relax
\EndOfBibitem
\bibitem[Mori \emph{et~al.}(2012)Mori, Glover, Schuurman, and Martinez]{Martinez_ethene}
T.~Mori, W.~J. Glover, M.~S. Schuurman and T.~J. Martinez, \emph{J. Phys. Chem. A}, 2012, \textbf{116}, 2808--2818\relax
\mciteBstWouldAddEndPuncttrue
\mciteSetBstMidEndSepPunct{\mcitedefaultmidpunct}
{\mcitedefaultendpunct}{\mcitedefaultseppunct}\relax
\EndOfBibitem
\bibitem[{Linstrom} and {Mallard}(retrieved 2021)]{nistwebn}
\emph{NIST Chemistry WebBook, NIST Standard Reference Database Number 69}, ed. P.~J. {Linstrom} and W.~G. {Mallard}, National Institute of Standards and Technology, Gaithersburg MD, 20899, retrieved 2021\relax
\mciteBstWouldAddEndPuncttrue
\mciteSetBstMidEndSepPunct{\mcitedefaultmidpunct}
{\mcitedefaultendpunct}{\mcitedefaultseppunct}\relax
\EndOfBibitem
\end{mcitethebibliography}
\bibliographystyle{rsc} 

\end{document}


\begin{titlepage}
	\begin{center}
    \vspace*{1in}
    	\LARGE{\textbf{Supplementary Information}}
        
        \vspace{0.5cm}
        \large{Henry G. McGhee, Henry J. Thompson, James Thompson, \\
            Yu Zhang, Adam S. Wyatt, Emma Springate, Richard T. Chapman, \\
            Rebecca A. Ingle, Daniel A. Horke, Russell S. Minns, Michael A. Parkes\\}
        
        \vspace{0.5cm}
        \small{\emph{Department of Chemistry, University College London, 20 Gordon Street, London, WC1H 0AJ, UK}}
        
        \vspace{0.5cm}
        E-mail: \href{michael.parkes@ucl.ac.uk}{michael.parkes@ucl.ac.uk}
    \end{center}
\end{titlepage}

\section{Calculated Geometry}

    \begin{table}[!ht]
    \centering
    \caption{Optimised structure for \emph{trans}-1,2-dichloroethene from a CCSD calculation with a 6-311G++(3df, 3pd) basis.}
    \begin{tabular}{|l|l|l|l|}
    \hline
        Atom & X / Bohr & Y / Bohr & Z / Bohr \\ \hline
        C & 0.493660842 & 0.442675934 & 0 \\ \hline
        H & 0.334309043 & 1.509306884 & 0 \\ \hline
        C & -0.493660842 & -0.442675934 & 0 \\ \hline
        H & -0.334309043 & -1.509306884 & 0 \\ \hline
        Cl & -2.144621671 & 0.048817983 & 0 \\ \hline
        Cl & 2.144621671 & -0.048817983 & 0 \\ \hline
    \end{tabular}
\end{table}

    \begin{table}[!ht]
    \centering
    \caption{Optimised structure for \emph{cis}-1,2-dichloroethene from a CCSD calculation with a 6-311G++(3df, 3pd) basis.}
    \begin{tabular}{|l|l|l|l|}
    \hline
        Atom & X / Bohr & Y / Bohr & Z / Bohr \\ \hline
        C & 0 & 0 & 0 \\ \hline
        H & 0 & 0 & 1.0792497 \\ \hline
        C & 1.1533616 & 0 & -0.6696234 \\ \hline
        Cl & -1.5402877 & 0 & -0.7320455 \\ \hline
        Cl & 1.2812994 & 0 & -2.3702141 \\ \hline
        H & 2.0906222 & 0 & -0.1345322 \\ \hline
    \end{tabular}
\end{table}
\newpage
\section{Vibrational Analysis}

    \begin{table}[!ht]
        \centering
        \caption{Calculated Frequencies for \emph{trans}-1,2-dichloroethene from a CCSD calculation with a 6-311G++(3df, 3pd) basis. Experimental values taken from NIST webbook \cite{nistwebn}}
        \begin{tabular}{|l|l|l|l|l|}
        \hline
            Symmetry & Calc / cm$^{-1}$ & NIST / cm$^{-1}$ & Period / fs & Character \\ \hline
            1 A$_u$ & 216.15 & 227 & 147 & Torsion \\ \hline
            2 B$_u$ & 243.09 & 250 & 133 & CCCl deform \\ \hline
            3 A$_g$ & 361.43 & 350 & 95 & CCCl deform \\ \hline
            4 B$_g$ & 798.56 & 763 & 44 & CH bend \\ \hline
            5 B$_u$ & 861.23 & 828 & 40 & CCl str \\ \hline
            6 A$_g$ & 886.05 & 846 & 39 & CCl str \\ \hline
            7 A$_u$ & 942.87 & 900 & 37 & CH bend \\ \hline
            8 B$_u$ & 1242.95 & 1200 & 28 & CH bend \\ \hline
            9 A$_g$ & 1318.17 & 1274 & 26 & CH bend \\ \hline
            10 A$_g$& 1677.01 & 1578 & 21 & CC str \\ \hline
            11 B$_u$& 3251.41 & 3073 & 11 & CH str \\ \hline
            12 A$_g$& 3253.76 & 3090 & 11 & CH str \\ \hline
        \end{tabular}
    \end{table}
    
    \begin{table}[!ht]
        \centering
        \caption{Calculated Frequencies for \emph{cis}-1,2-dichloroethene from a CCSD calculation with a 6-311G++(3df, 3pd) basis. Experimental values taken from NIST webbook \cite{nistwebn}}
        \begin{tabular}{|l|l|l|l|l|}
            \hline
            Symmetry & Calc / cm$^{-1}$ & NIST / cm$^{-1}$ & Period / fs & Character \\ \hline
            1 A$_1$& 172.56 & 173 & 193 & CCCl deform \\ \hline
            2 A$_2$& 421.04 & 406 & 82 & Torsion \\ \hline
            3 B$_1$& 589.81 & 571 & 58 & CCCl deform \\ \hline
            4 B$_2$& 720.82 & 697 & 48 & CH bend \\ \hline
            5 A$_1$& 760.19 & 711 & 47 & CCl str \\ \hline
            6 B$_1$& 902.92 & 857 & 39 & CCl str \\ \hline
            7 A$_2$& 905.75 & 876 & 38 & CH bend \\ \hline
            8 A$_1$& 1233.8 & 1179 & 28 & CH bend \\ \hline
            9 B$_1$& 1343.76 & 1303 & 25 & CH bend \\ \hline
            10 A$_1$& 1662.24 & 1587 & 21 & CC str \\ \hline
            11 B$_1$& 3229.88 & 3072 & 11 & CH str \\ \hline
            12 A$_1$& 3252.34 & 3077 & 11 & CH str \\ \hline
        \end{tabular}
    \end{table}

    \begin{figure}[!ht]
        \centering
        \includegraphics{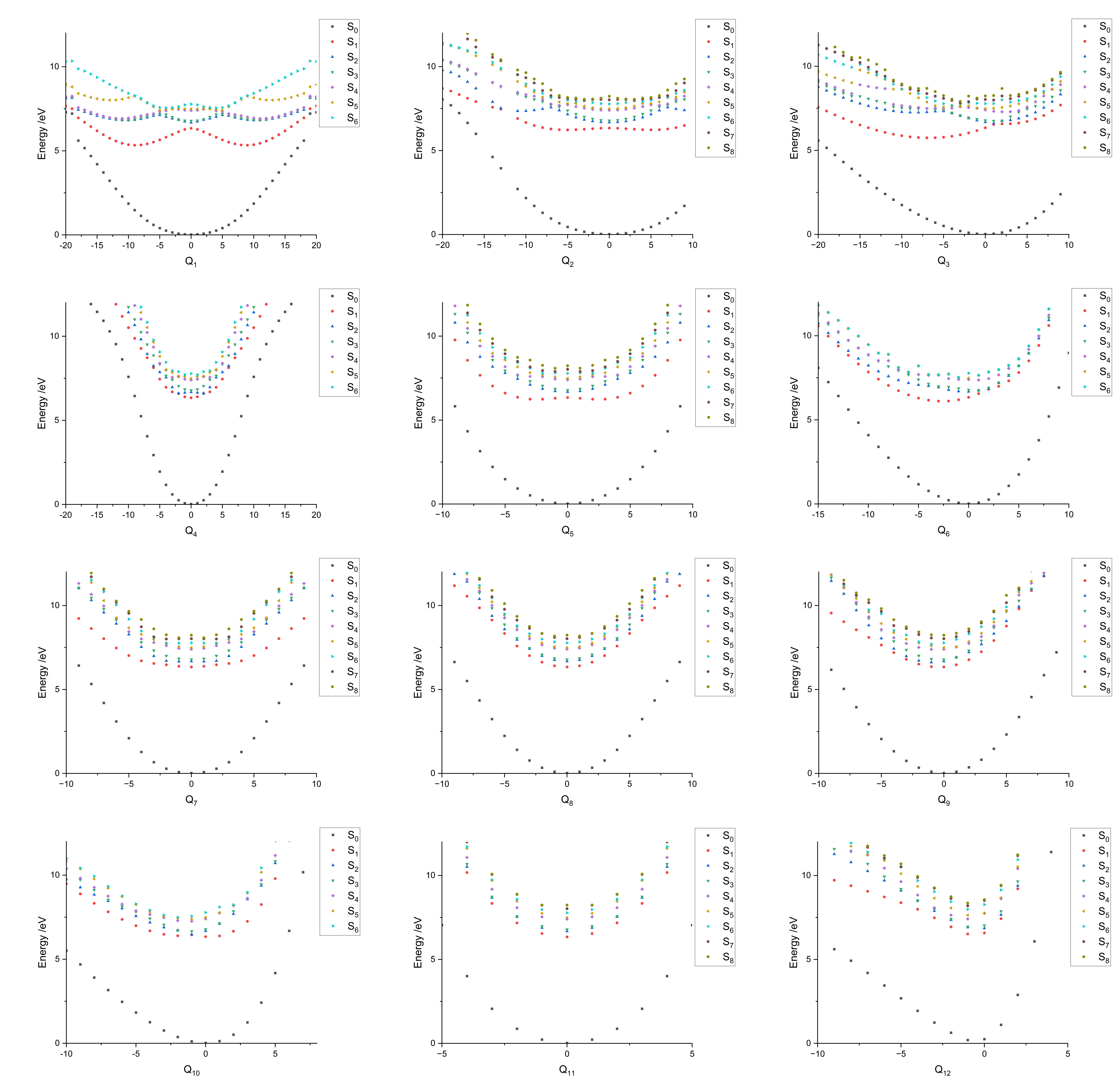}
        \caption{Cuts through the potential energy surface of \emph{trans}-1,2-dichloroethene along mass-weighted normal modes. Calculated using EOM-CCSD/6-311G++(3df,3pd) in QChem. The number of the mode is indicated in the x-axis label.}
        \label{fig:si_surfaces}
    \end{figure}
\newpage

\section{Ionization Energies}

    \begin{figure}[!ht]
        \centering
        \includegraphics[width=12cm]{transPES_comp.png}
        \caption{Comparison of our \emph{trans}-1,2-dichloroethene  XUV photoelectron spectrum at 21.64~eV with calculated ionization energies using EOM-CCSD with a 6-311G++(3df, 3pd) basis set}
        \label{fig:si_transpes}
    \end{figure}

    \begin{figure}[!ht]
        \centering
        \includegraphics[width=12cm]{cispescomp.png}
        \caption{Comparison of our \emph{cis}-1,2-dichloroethene XUV photoelectron spectrum at 21.64~eV with calculated ionization energies using EOM-CCSD with a 6-311G++(3df, 3pd) basis set}
        \label{fig:si_cispes}
    \end{figure}
\newpage

\section{Photoabsorption}

    \begin{figure}[!ht]
        \centering
        \includegraphics[width=12cm]{trans_PA.png}
        \caption{UV-vis spectrum of \emph{trans}-1,2-dichloroethene taken from Locht \emph{et al},\cite{Locht2020-trans} shown with position of pump pulse (black Gaussian) and the calculated positions of the first three excited states from EOM-CCSD calculations. The green lines are for states with no or very low oscillator strengths}
        \label{fig:si_transpa}
    \end{figure}

    \begin{figure}[!ht]
        \centering
        \includegraphics[width=12cm]{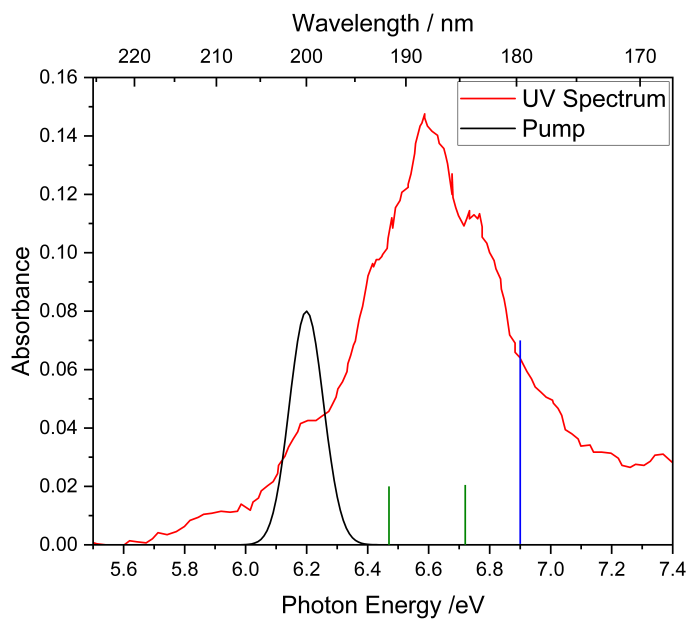}
        \caption{UV-vis spectrum of \emph{cis}-1,2-dichloroethene taken from Locht \emph{et al},\cite{Locht2019-cis} shown with position of pump pulse (black Gaussian) and the calculated positions of the first three excited states from EOM-CCSD calculations. The green lines are for states with no or very low oscillator strengths}
        \label{fig:si_cispa}
    \end{figure}
\newpage

\section{\emph{Cis}-1,2-dichloroethene time resolved photoelectron spectroscopy results}

    \begin{figure}[!ht]
        \centering
        \includegraphics{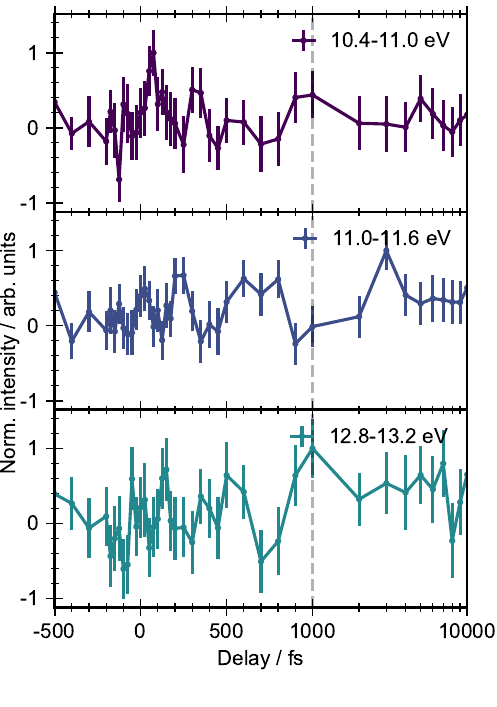}
        \caption{Integrated kinetic traces for the expected photoproduct regions (10.4-11.0, 11.0-11.6 eV and 12.8-13.2 eV) for \emph{cis}-1,2-dichloroethene. No signal is observed above the noise level.}
        \label{fig:cis_photoproducts}
    \end{figure}

\newpage
\section{Bleach Signals}
    \begin{figure}[!ht]
        \centering
        \begin{subfigure}[b]{0.45\textwidth}
            \centering
            \includegraphics[width=\textwidth]{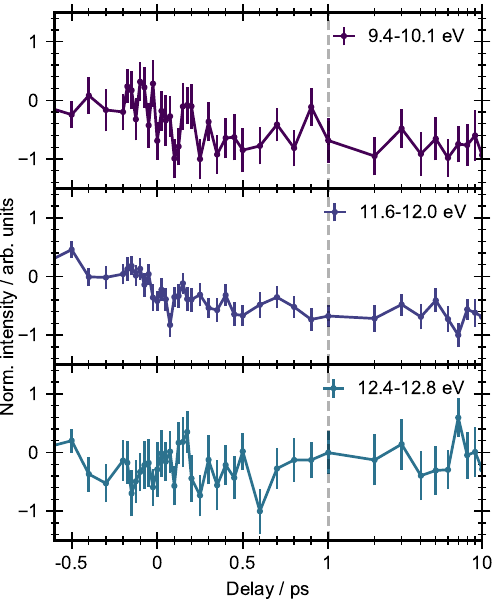}
            \caption{}
            \label{fig:trans_bleach}
        \end{subfigure}
        \hfill
        \begin{subfigure}[b]{0.45\textwidth}
            \centering
            \includegraphics[width=\textwidth]{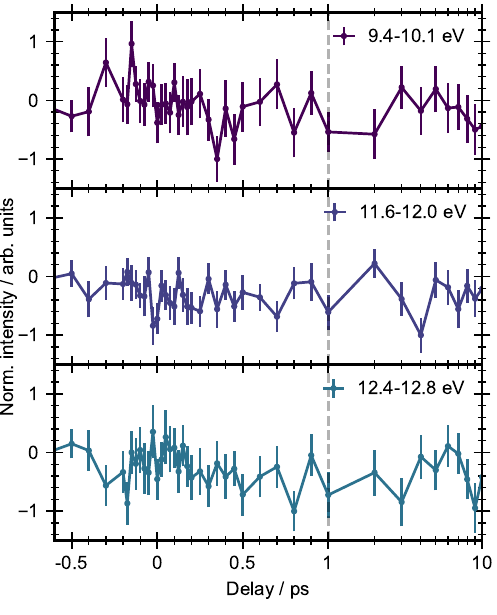}
            \caption{}
            \label{fig:cis_bleach}
        \end{subfigure}
        \caption{Integrated kinetic traces of the ground state bleach features for the first (9.4-10.1~eV), second (11.6-12.0~eV) and third (12.4-12.8~eV) ionisation potentials for (a) \emph{trans}-1,2-dichloroethene and (b) \emph{cis}-1,2-dichloroethene. The delay axis is plotted on a log scale for values greater than 1~ps.}
        \label{fig:bleach_signals}
    \end{figure}

\newpage
\section{Cl region}

\begin{figure}[h!]
    \centering
    \includegraphics{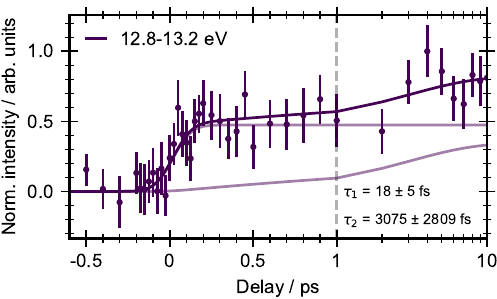}
    \caption{Integrated kinetic trace of the Cl region (12.8-13.2~eV) showing the fit with the sum of two exponential rises (convoluted with the Gaussian IRF). Individual fit components are shown by faint lines and the sum is shown by the bold line. The longer time constant may be concurrent with the slow decay ($\sim$2~ps) of the 8.8-9.2~eV region, however the errors on the fit are large and so no definitive assignments can be made.}
    \label{fig:Cl_long_delay}
\end{figure}

\section{\ce{C2HCl} region}

\begin{figure}[h!]
    \centering
    \includegraphics{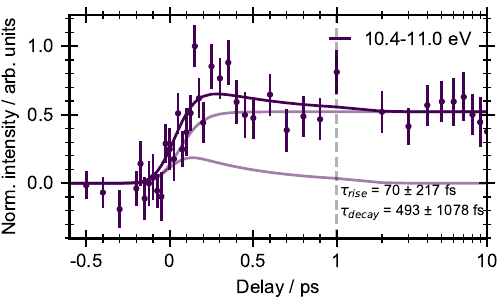}
    \caption{Integrated kinetic trace for the \ce{C2HCl} region (10.4-11.0~eV) showing the fit with an exponential decay and an exponential rise in order to account for the peak in signal at around 0.3~ps. While this qualitatively produces a better fit, the errors from the least-squares fitting process are very large and so no definitive assignments can be made.}
    \label{fig:C2HCl}
\end{figure}

\newpage

\bibliography{Article} 
\bibliographystyle{rsc} 